\documentclass[journal,twoside,confmode]{IEEEtran} 
\usepackage[noadjust]{cite}
\usepackage{tmi}
\usepackage{cite}
\usepackage{amsmath,amssymb,amsfonts}
\usepackage{algorithmic}
\usepackage{graphicx}
\usepackage{textcomp}
\usepackage{multirow}
\usepackage{url}
\usepackage{diagbox}
\usepackage{subcaption}
\usepackage{hyperref}
\usepackage{soul}
\newcommand\itblue[1]{\textcolor{black}{{#1}}}
\def\BibTeX{{\rm B\kern-.05em{\sc i\kern-.025em b}\kern-.08em
    T\kern-.1667em\lower.7ex\hbox{E}\kern-.125emX}}
\begin{document}
\title{Segmentation-Renormalized Deep Feature Modulation for Unpaired Image Harmonization}
\author{Mengwei Ren\textsuperscript{*}, Neel Dey\textsuperscript{*}, James Fishbaugh, Guido Gerig, \IEEEmembership{Fellow, IEEE}
\thanks{The authors were supported by NIH grants 1R01DA038215-01A1 (PI Grewen), R01-HD055741-12 (PI Piven), 1R01HD088125-01A1 (PI Botteron), 1R01MH118362-01 (PI Pruett), 2R01EY013178-15 (PI Schumann).}
\thanks{* These authors contributed equally. All authors are with the Department of Computer Science and Engineering at New York University Tandon School of Engineering, Brooklyn, NY 11201 USA (e-mail addresses: \{mengwei.ren, neel.dey, james.fishbaugh, gerig\}@nyu.edu).
}}

\maketitle

\begin{abstract}
Deep networks are now ubiquitous in large-scale multi-center imaging studies. However, the direct aggregation of images across sites is contraindicated for downstream statistical and deep learning-based image analysis due to inconsistent contrast, resolution, and noise. To this end, \itblue{in the absence of paired data}, variations of Cycle-consistent Generative Adversarial Networks have been used to harmonize image sets between a source and target domain. \itblue{Importantly}, these methods are prone to instability, contrast inversion, intractable manipulation of pathology, and steganographic mappings which limit their reliable adoption in real-world medical imaging. In this work, based on an underlying assumption that morphological \itblue{shape} is consistent across imaging sites, 
we propose a \itblue{segmentation-renormalized image translation framework} to reduce inter-scanner heterogeneity while preserving anatomical layout.
We replace the affine transformations used in the normalization layers within generative networks with trainable scale and shift parameters conditioned on jointly learned anatomical segmentation \itblue{embeddings} to modulate features at every level of translation. We evaluate our methodologies against recent baselines across several imaging modalities (T1w MRI, FLAIR MRI, and OCT) on datasets with and without lesions. Segmentation-renormalization for translation GANs yields superior image harmonization as quantified by Inception distances, demonstrates improved downstream utility via post-hoc segmentation accuracy, and improved robustness to translation perturbation and self-adversarial attacks.
\end{abstract}


\section{Introduction}
\label{sec:introduction}
\begin{figure}[t]
    \centering
    \includegraphics[width=0.47\textwidth]{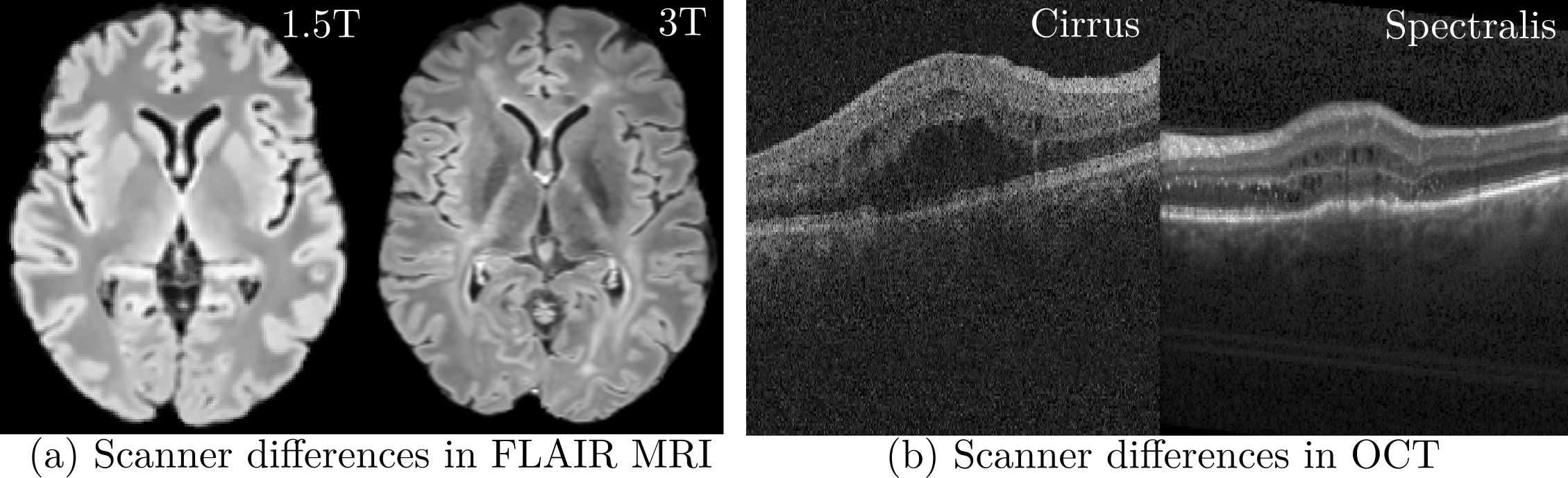}
    \caption{\itblue{Unpaired samples across medical imaging scanners illustrating inter-device appearance variability.}}
    \label{fig:example}
\end{figure}

Large-scale multi-center imaging studies acquire data over several years and analyze them jointly to track potential bio-markers for specific diseases. However, data collection may involve the upgrade of imaging devices over time or the use of multiple scanners in parallel, introducing non-biological variability and batch-effects into any statistical analysis. For example, magnetic resonance images acquired from $1.5$ T and $3$ T scanners differ significantly \itblue{in intensity, contrast, and noise distribution (see Fig.~\ref{fig:example} (a)), with even higher inter-scanner variability observed in Optical Coherence Tomography (OCT) as in Fig.~\ref{fig:example} (b).}
Therefore, harmonization to a consistent \itblue{imaging} standard is critical to downstream tasks. Further, harmonization enables the introduction of legacy data into new studies, thereby accelerating collaborative imaging research by increasing statistical power and reducing costs.

The multi-site imaging harmonization problem has been studied extensively, with varying approaches \cite{POMPONIO2020116450,GARCIADIAS2020117127,blumberg2019}. One strategy takes an Empirical Bayes approach \cite{10.1093/biostatistics/kxj037} to harmonize image-derived statistical values such as cortical thickness across sites to an intermediate domain between source and target. Another image-specific approach attempts to normalize intensity between scanners via non-parametric density flows \cite{FORTIN2018104}. Given paired images, i.e. the same subject scanned on different devices, supervised learning with convolutional networks can be applied to learn mappings between medical imaging domains~\cite{10.1007/978-3-030-00536-8_3,DEWEY2019160}. However, acquiring paired images is expensive and logistically challenging for most study designs.

Recently, Cycle-consistent Generative Adversarial Networks~\cite{Zhu_2017_ICCV} (CycleGAN) and its derivatives have been successful in unpaired image-to-image translation with direct applicability to the unpaired harmonization task \itblue{when the imaging sites have roughly similar subject demographics}. For example,~\cite{8759158} applies CycleGAN to OCT harmonization and~\cite{10.1007/978-3-030-32251-9_52} uses spherical U-Nets within CycleGAN for cortical thickness harmonization. However, cycle-consistent adversarial methods are prone to instability, can invert contrast, and can manipulate or introduce pathological lesions \cite{zhang2018harmonic,cohen2018distribution}. 
Further, as translation between domains with differing amounts of structural information is ill-posed (as in image harmonization), a cycle-consistent generator may employ \textit{self}-adversarial noise and learn mappings that are highly susceptible to noise \cite{chu2017cyclegan,bashkirova2019adversarial} making its outputs unreliable for downstream tasks.

\begin{figure*}[!ht]
\centering
\includegraphics[width=\textwidth]{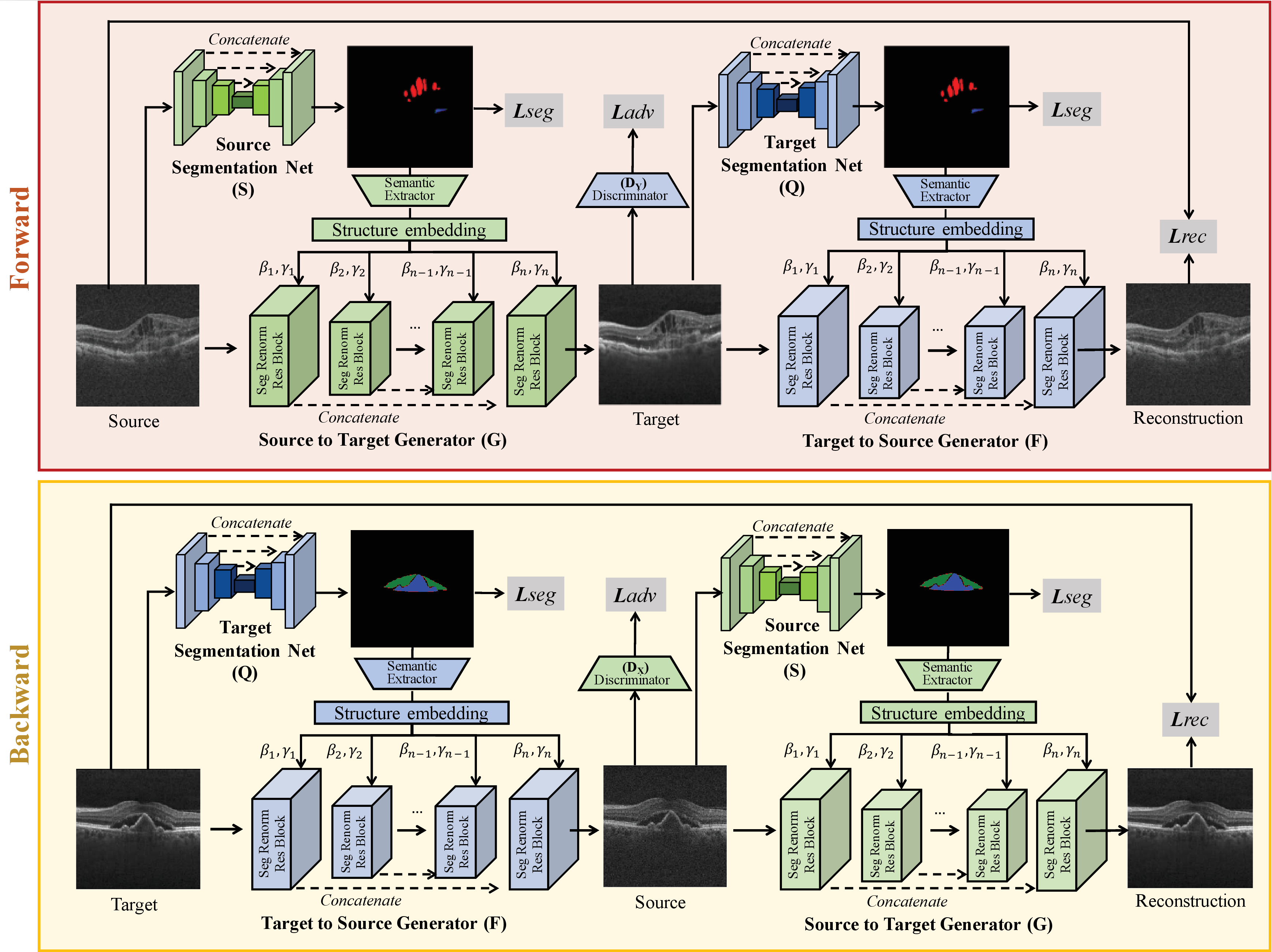}
   \caption{A high-level overview of the forward\itblue{/backward} translation in our \itblue{framework}. \itblue{The generator learns} cross-domain translations with \itblue{its normalized} layer-wise feature-maps \itblue{affinely} modulated by a jointly learned shared segmentation-embedding \itblue{derived from the semantic extractor subnetwork}. \itblue{\textbf{Seg Renorm Res Block} denotes the proposed segmentation-renormalized residual block detailed in Figure \ref{fig:seg-aware-norm}. Green subnetworks are source domain-specific, blue subnetworks are target domain-specific, and gray boxes correspond to our objective functions.}
   }
\label{fig:pipeline}
\end{figure*}

Current work has sought to improve translation quality and robustness via multi-task strategies. For instance, translation can be improved by imposing a segmentation consistency loss between input and output~\cite{Zhang_2018_CVPR, DBLP:journals/corr/abs-1807-04409}, or by further providing an instance mask coupled with the input to the generator~\cite{mo2018instanceaware}. In contrast to the goal of leveraging segmentation for high-quality robust image translation, another line of work proposes to segment target modalities \textit{without} labels by utilizing translation GANs and source modality labels~\cite{huo2018adversarial, chen2020unsupervised}. While these methods regularize translation at the image (input/output) level, they do not consider \textit{feature}-level regularization.

Common GAN generators are constructed as stacks of repeating convolutions, normalizations, and activations. Unfortunately, the normalization layers can `wash away' spatial context~\cite{DBLP:journals/corr/abs-1903-07291} \itblue{and} may be detrimental to harmonization given that anatomical structure should remain consistent between domains. Feature-level segmentation-conditioning for \textit{paired} translation was proposed in \cite{DBLP:journals/corr/abs-1903-07291}. However, it learned affine parameters for each index in a feature tensor and thus does not scale to \textit{unpaired} problems which require a reverse translation.

In this work, we improve unpaired image-to-image translation and hence image harmonization by conditionally normalizing cycle-consistent adversarial methods at the feature level via segmentation-derived linear modulation ~\cite{DBLP:journals/corr/HuangB17, perez:hal-01648685}. In the normalization layers of the generator, we first normalize features to zero mean and unit standard deviation, and then renormalize them with affine scale and shift transformations whose parameters are learned from a jointly trained semantic segmentation\itblue{-derived embedding} branch.  

Importantly, our approach is generic and could be applied to any other image translation framework where segmentation labels are available. Our contributions are as follows,
\begin{itemize}
    \item We improve on cycle-consistent adversarial methods in the unpaired multi-site harmonization task with a novel segmentation-aware renormalization layer;
    \item Our proposed method regularizes mappings towards maintaining anatomical segmentation consistency while eliminating non-biological batch-effects;
    \item We evaluate the proposed methodologies against both standard and segmentation-aware baselines across diverse imaging modalities (T1w MRI, FLAIR MRI, and OCT) via sample fidelity, post-hoc segmentation accuracy scores, and sensitivity to translation perturbation.
    \item Finally, we demonstrate the robustness of our methods in harmonization tasks both in the context of pathological image translation where standard methods hallucinate lesions (or lack thereof), and in robustness to adversarial attacks in image translations.
\end{itemize}
\itblue{Our image translation code is publicly available\footnote{\url{https://github.com/mengweiren/segmentation-renormalized-harmonization}}}.

\section{Methods}

\subsection{Methodological Overview}
We formalize unpaired multi-site harmonization \itblue{as an image-to-image translation} problem between two domains $X$ and $Y$. An overview of the proposed pipeline is shown in Fig.~\ref{fig:pipeline}. 
In canonical cycle-consistent translation methods, the forward mapping is defined as $G: X \rightarrow Y$, harmonizing images \itblue{from source scanner $X$ to target scanner $Y$}, such that $G(X)$ is indistinguishable from $Y$ to the discriminator $D_Y$. As $G$ is learned from unpaired samples
, the framework additionally learns an inverse mapping $F: Y \rightarrow X$ and enforces a cycle-consistency constraint $F(G(X)) \approx X$ to achieve unpaired translation consistent with the input.

We begin \itblue{our improvements} by incorporating segmentation networks $S$ and $Q$ in the CycleGAN setting \itblue{as shown in Fig.~\ref{fig:pipeline}}, where $S$ is trained with (image, segmentation)-pairs
from the source domain, and $Q$ is \itblue{trained with}
pairs \itblue{from} the target domain. \itblue{Motivated by the assumption that morphological shape is consistent across imaging domains, $S$ and $Q$ are encouraged to produce the same segmentation via a segmentation-objective as in \cite{Zhang_2018_CVPR,DBLP:journals/corr/abs-1807-04409}.}

In addition to a loss-based approach, we \itblue{further} use segmentation information to adaptively control the scales and biases of each layer within the generator. 
\itblue{Using a semantic extractor subnetwork (shown in Fig.~\ref{fig:pipeline}), we derive a learned structural embedding from the predicted segmentation map which is then used to channel-wise condition the normalized output of each convolutional layer} of the generator via Feature-wise Linear Modulation (FiLM)~\cite{perez:hal-01648685}. \itblue{We hypothesize that feature-level segmentation-conditioning provides improved translation generator activations over using input/output-level segmentation-regularization alone.  Importantly}, as the structural embedding is learned using a convolutional network (which are in practice neither translation-equivariant \cite{kayhan2020translation,zhang19shift} \itblue{nor deformation-equivariant}), the embedding still contains information about the relative positioning of anatomical structures.

\subsection{Segmentation-renormalization.} 
\begin{figure}[t]
    \centering
    \includegraphics[width=0.45\textwidth]{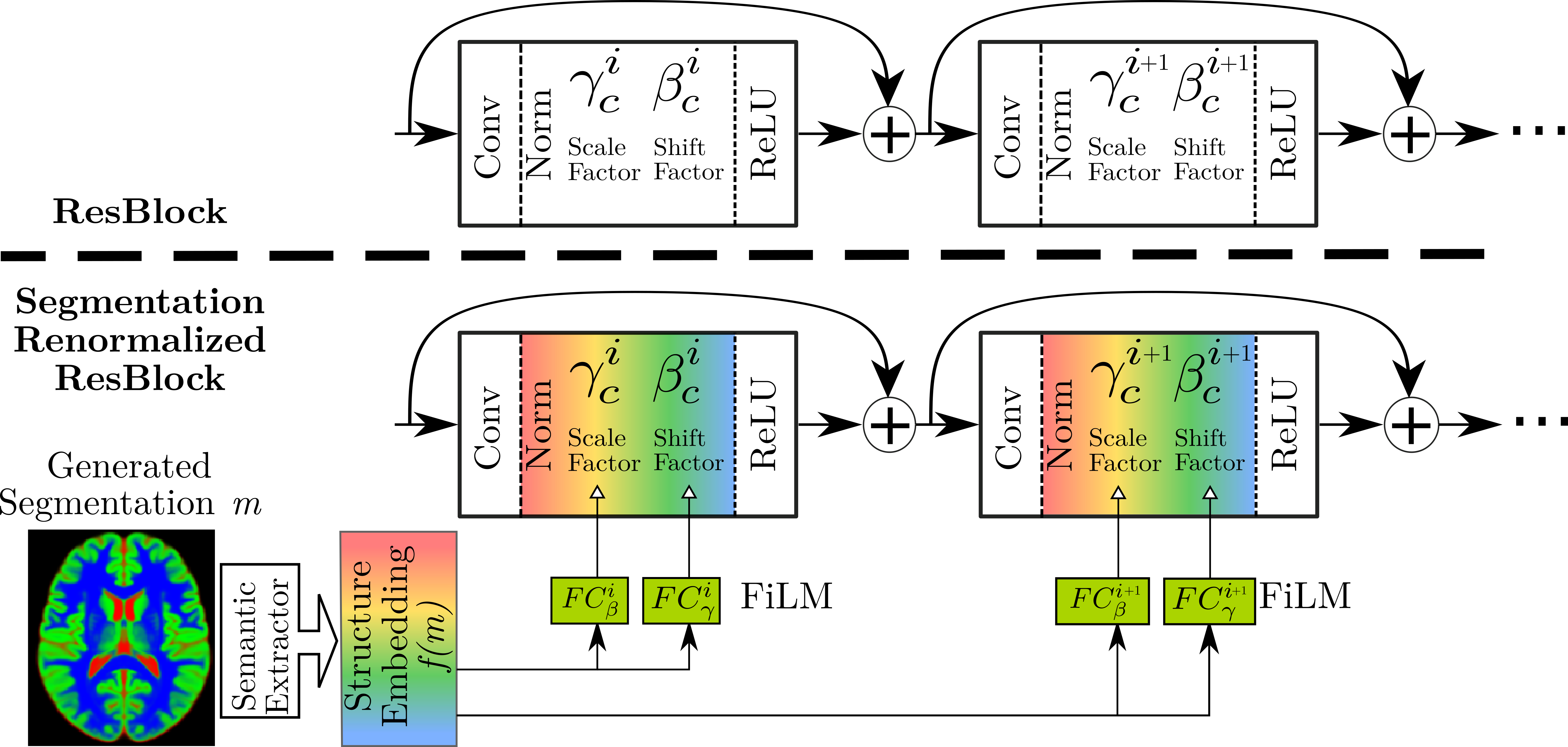}
    \caption{A comparison between (\textbf{top}) standard residual blocks and (\textbf{bottom}) our proposed segmentation-renormalized residual blocks \itblue{(Seg Renorm Res Block)}.}
    \label{fig:seg-aware-norm}
\end{figure}

We incorporate anatomical priors \itblue{by replacing the convolutional residual blocks in the generator subnetworks with the proposed segmentation-renormalized residual block as illustrated in Fig.~\ref{fig:seg-aware-norm}}, modulating intermediate features with an affine transformation conditioned on a learned segmentation.  
Formally, a standardization and renormalization is inserted into each residual block of generators $G$ and $F$ between the convolutional layers and their activations. Features in the generator are standardized in a channel-wise manner and linearly modulated with learnable scales and shifts defined as,
\begin{equation}
    \gamma_{c}^{i} \Big( \frac{h_{n, c}^{i}-\mu_{n,c}^{i}}{\sigma_{n,c}^{i}} \Big) + \beta_{c}^{i},
\end{equation}
where $h_{n, c}^i$ is the \itblue{2D spatial} output of the convolutional layer and $\mu_{n,c}^{i}$ and $\sigma_{n,c}^{i}$ are the mean and standard deviation of channel $c$ in the $i$th block from the $n$th sample in the batch. Fig.~\ref{fig:seg-aware-norm} shows a comparison between a standard residual block~\cite{He_2016_CVPR} and the proposed residual block.
$\gamma_{c}^{i}$ and $\beta_{c}^{i}$ are renormalization parameters that scale and shift the normalized feature at each channel. In contrast to standard normalization, we learn $\gamma_{c}^{i}$ and $\beta_{c}^{i}$ \itblue{using FiLM layers (i.e., two fully connected layers $FC^i_\gamma$ and $FC^i_\beta$ in our implementation) on the  structure embedding $f(m)$ (which is derived from a predicted segmentation mask $m$) such that: $\gamma_{c}^{i} = FC^i_\gamma(f(m))$, and $\beta_{c}^{i} = FC^i_\beta(f\mathbf(m))$. }To enable feature reuse, this structural embedding is shared across all blocks \itblue{as shown in Figs.~\ref{fig:pipeline} and \ref{fig:net_generator}} and is then linearly modulated for each \itblue{translation layer} separately.

\itblue{We note that conditional normalization has generically been extensively studied for both image translation~\cite{DBLP:journals/corr/HuangB17,DBLP:journals/corr/abs-1903-07291} and medical imaging segmentation~\cite{chartsiasdisentangle,jacenkow2020inside}. We instead propose to conditionally normalize generative unpaired translation networks based on learned segmentation embeddings, leading to both improved translation fidelity and downstream utility as shown in Section \ref{sec:experiments}.}

\subsection{Architecture and Design Details}
The proposed framework \itblue{as shown in Fig. \ref{fig:pipeline}} can be decomposed \itblue{into} segmentation networks, \itblue{semantic extractors}, generators, and discriminators.
\begin{figure}
    \centering
    \includegraphics[width=0.45\textwidth]{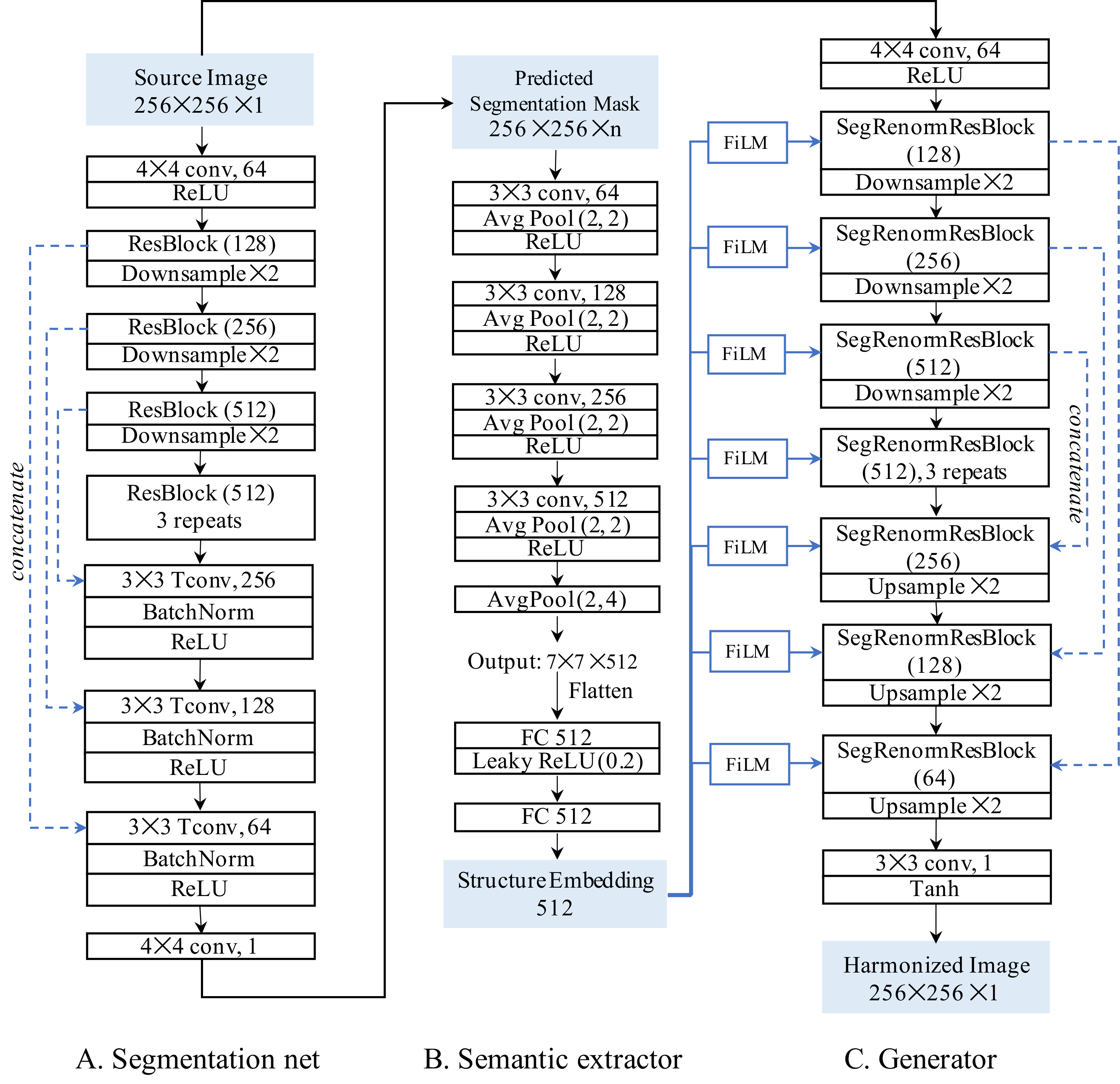}
    \caption{\itblue{Overall translator architecture with subnetworks \textbf{A.} segmentation net, \textbf{B.} semantic extractor, and \textbf{C.} segmentation-renormalized generator. $n$ is the number of classes in the segmentation. For the IXI experiments, input spatial resolutions are $128\times 128$ and the \textit{AvgPool}$(2,4)$ in B. was replaced by \textit{AvgPool}$(1,2)$.}}
    \label{fig:net_generator}
\end{figure}
\begin{figure}[t]
    \centering
    \includegraphics[width=0.12\textwidth]{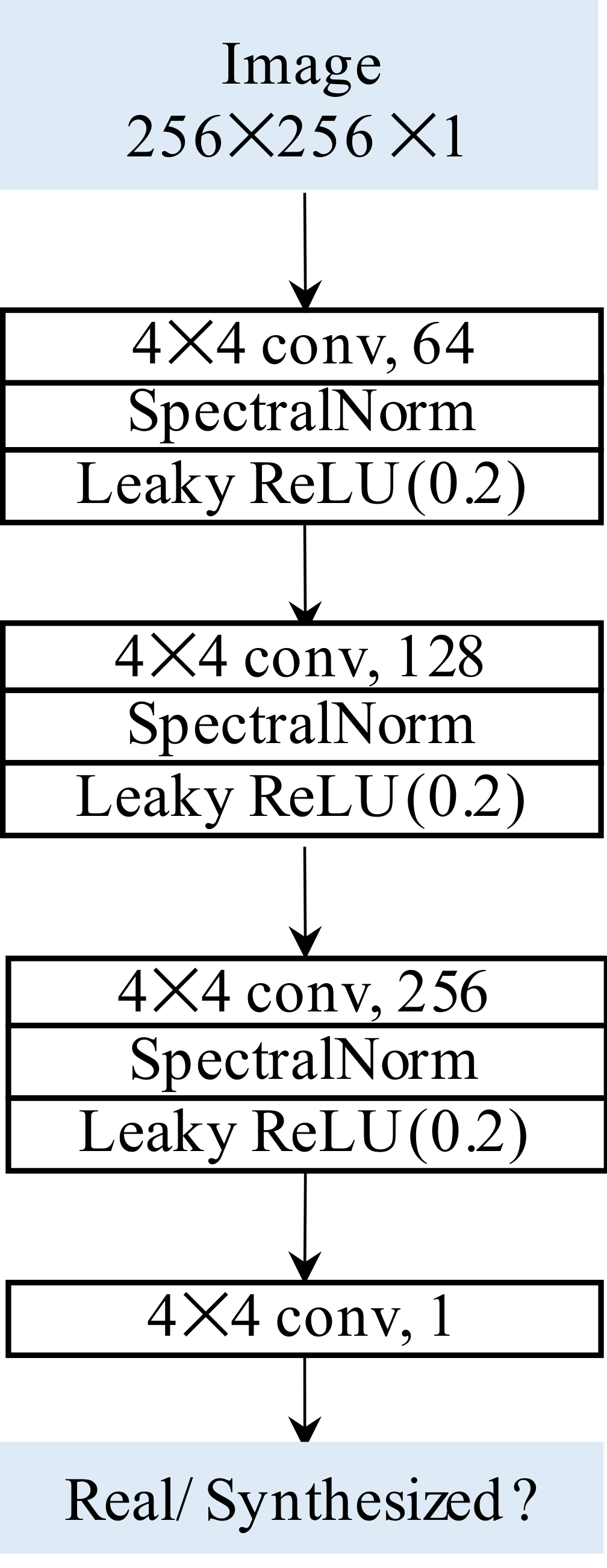}
    \caption{\itblue{Patch discriminator network architecture. For the IXI experiments, the input image resolution is $128\times 128\times1$.}}
    \label{fig:net_discriminator}
\end{figure}
\itblue{Briefly, the segmentation network, semantic extractor, and generator jointly form the image-to-image translator with the discriminator trained in combination to calculate the adversarial objective. \itblue{Architectural details are given in Figs.~\ref{fig:net_generator} and~\ref{fig:net_discriminator} using the following nomenclature:}
 \begin{itemize}
    \item \underline{ResBlock(k)} represents two repeating \{k $3\times3$ convolutions, batch normalization, ReLU\} blocks with an additive skip connection as illustrated in Fig.~\ref{fig:seg-aware-norm}.
    \item Analogously, \underline{SegRenormResBlock(k)} denotes the proposed residual block where each normalization is replaced by the segmentation-renormalization illustrated in Fig.~\ref{fig:seg-aware-norm}.
    \item \underline{TConv} represents a transposed convolutional layer.
    \item Lastly, \underline{AvgPool(k,s)} denotes an average pooling operation with kernel size $k$ and stride size $s$.
\end{itemize}}
\subsubsection{Segmentation nets} U-Nets~\cite{DBLP:journals/corr/RonnebergerFB15} \itblue{detailed in Fig. \ref{fig:net_generator} A are used} to produce segmentation masks from input images to \itblue{both} regularize the image translation \itblue{via segmentation-consistency and to provide an input to the semantic extractor, detailed below}. These auxiliary networks are trained under a weighted cross-entropy cost for its robustness to label noise induced by imperfect segmentation~\cite{rolnick2017deep}.

\subsubsection{Semantic Extractors} \itblue{Taking the predicted segmentation as input, a shallow CNN (detailed in Fig.~\ref{fig:net_generator} B) extracts a semantic embedding that serves as input to FiLM layers which modulate generator features as described below.}

\subsubsection{Generators}  Residual U-Nets \itblue{(detailed in Fig. \ref{fig:net_generator} C)} are used for the generator networks with its residual blocks replaced with our proposed segmentation-renormalized residual blocks. The proposed renormalization is used throughout the generator to propagate the learnt segmentation embedding at multiple resolutions during synthesis.

\subsubsection{Discriminators} We adopt PatchGAN discriminators~\cite{Zhu_2017_ICCV} as described in \itblue{Fig. \ref{fig:net_discriminator}}, distinguishing between \itblue{real and synthesized images} at the patch level with patch size determined by the network receptive field ($34\times 34$ in our implementation). \itblue{Spectral normalization~\cite{DBLP:journals/corr/abs-1802-05957} was used in the discriminator for training stability}. 

\subsection{Learning objectives}
The framework is trained end-to-end with multiple losses.

\subsubsection{Adversarial terms} We employ a least squares objective for adversarial training \cite{mao2017least}, due to its improved stability over a cross-entropy objective in our task. The two-player adversarial game is optimized as,
\begin{align}
    \begin{split}
    \min_{D_Y}\mathcal{L}_{\mathrm{GAN}}(D_Y) &= \frac{1}{2}\mathbb{E}_{y \sim Y}\left[||1-D_Y(y)||^2_2\right] \\
    &+ \frac{1}{2}\mathbb{E}_{x \sim X}\left[||D_Y(G(x))||^2_2\right],\\
    \min_G\mathcal{L}_{\mathrm{GAN}}(G) &= \frac{1}{2}\mathbb{E}_{x\sim X}\left[||D_Y(G(x)) -1 ||_2^2\right],
    \end{split}
\end{align}
with analogous optimization applied for $F$ and $D_X$.

\subsubsection{Segmentation terms}

\itblue{The segmentation networks S and Q shown in Fig. \ref{fig:pipeline} for source and target domains, respectively, are trained under a weighted cross-entropy objective. In the forward cycle, the source domain sample $x$ has a groundtruth segmentation mask which is used as a reference for subnetwork $S$. However, once $x$ is translated to the target domain, $G(x)$ does not have expert annotation. Motivated by our assumption that anatomical layout is consistent across domains, we give the segmentation subnetwork $Q$ the same segmentation reference as $S$ (the groundtruth source domain labels). Analogous reasoning follows for the backward cycle, where subnetwork $S$ is given the same reference as $Q$ (the groundtruth target domain labels). The loss is defined as,
\begin{align}
\label{eq:seg_loss}
    \begin{split}
        \mathcal{L}_{\mathrm{seg}}(S, Q) = \mathbb{E}_{x \sim X}[ &-\sum_{c=0}^{n-1}\lambda_cs_{c}\log(S(x,c)) \\
        &-\sum_{c=0}^{n-1}\lambda_cs_{c}\log(Q(G(x),c))] \\
        +\mathbb{E}_{y \sim Y}[ &-\sum_{c=0}^{n-1}\lambda_cq_{c}\log(Q(y,c)) \\
        &-\sum_{c=0}^{n-1}\lambda_cq_{c}\log(S(F(y),c))],
    \end{split}
\end{align}
where $n$ is the number of classes, $S(x, c)$ is the prediction for class $c$ from input $x$, $s_c$ is the ground truth mask acquired from source domain,  $\lambda_0$ indicates the weights applied on negative samples (background), and $\lambda_1, \dots, \lambda_{n-1}$ are weights for positive samples (foreground). $Q(y,c)$ and $q_c$ are analogously defined in the target domain.}

\subsubsection{Cycle-consistency terms} 
To reduce the space of possible mapping functions and enable training with an unpaired dataset, a cycle consistency loss is defined as,
\begin{align}
    \begin{split}
    \mathcal{L}_{\mathrm{cyc}}(G, F) &=\mathbb{E}_{x \sim X}\left[\|F(G(x))-x\|_{1}\right] \\
    &+\mathbb{E}_{y \sim Y}\left[\|G(F(y))-y\|_{1}\right].
    \end{split}
\end{align}

\subsubsection{Total objective} The \itblue{complete} objective function of our model to minimize is summarized as,
\begin{align}
    \begin{split}
    &\mathcal{L}(G, F, D_X, D_Y, S, Q) = \\ &\lambda_{GAN}[\mathcal{L}_{\mathrm{GAN}}(G) + \mathcal{L}_{\mathrm{GAN}}(D_Y) + \mathcal{L}_{\mathrm{GAN}}(F) + \mathcal{L}_{\mathrm{GAN}}(D_X)] + \\ &\lambda_{cyc}\mathcal{L}_{\mathrm{cyc}}(G, F) \ + \lambda_{seg}\mathcal{L}_{\mathrm{seg}}(S,Q).
    \end{split}
\end{align} 

\begin{figure*}[!ht]
\begin{center}
\includegraphics[width=1.0\textwidth]{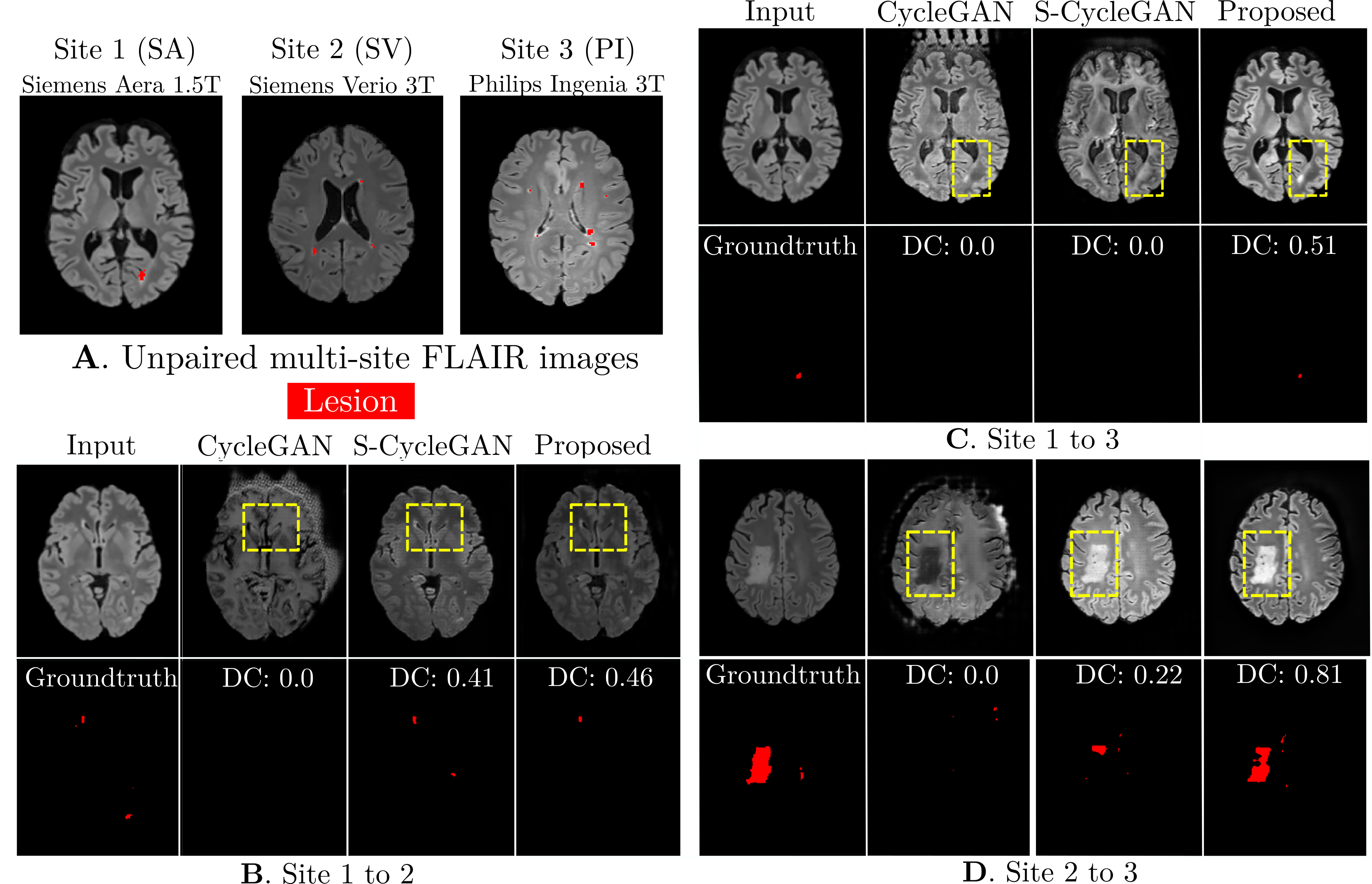}
\end{center}
   \caption{Harmonization results on test 2D multi-site FLAIR slices from MS-SEG. \itblue{Downstream} segmentation results are shown below \itblue{translated} images, annotated with Dice Coefficients (DC) for the visualized slices. \textbf{A.} Example slices from three sites, highlighting their varying contrasts; \textbf{B.} SA $\rightarrow$ SV translation; \textbf{C.} SV $\rightarrow$ PI translation; \textbf{D.} SA $\rightarrow$ PI translation. \itblue{Severe prediction artefacts appear in baseline outputs whereas the proposed model preserves semantic layout and appearance. For example, note the contrast inversion of tissue in \textbf{B} and the lesion in \textbf{D} (see yellow insets) using CycleGAN and overall contrast inversion in \textbf{C} using S-CycleGAN. Strong decreases in downstream segmentation performance appear across baselines with false-positive and false-negative examples marked by white and yellow arrows, respectively. In all settings, the proposed methodology demonstrates significant improvements in both translation fidelity and post-hoc segmentation performance. 
   }}
\label{fig:result-msseg}
\end{figure*}

\section{Experiments} \label{sec:experiments}
\subsection{Data and preprocessing:} We tested our method across several modalities: MRI (T1w, FLAIR) and OCT on three public datasets \itblue{IXI~\cite{IXI}, MS-SEG~\cite{commowick_objective_2018}, and RETOUCH~\cite{RETOUCH}, respectively}. \itblue{For IXI and RETOUCH,} we used a 70/30 train/test split at the subject-level. \itblue{For MS-SEG we performed leave-one-subject-out cross-validation as only five subjects were imaged per scanner.} In all datasets, no individual was scanned on more than one device.  

\subsubsection{MS-SEG} FLAIR MRI of subjects with Multiple Sclerosis (MS) were collected for an MS lesion segmentation challenge \cite{commowick_objective_2018}. Five non-overlapping subjects per scanner were imaged on three different MR scanners: Siemens Aera $1.5$ T (SA), Philips Ingenia $3$ T (PI), and Siemens Verio $3$ T (SV). The scanner specifications are detailed in Table~\ref{tab:msseg-scanner}. \itblue{See Fig.~\ref{fig:result-msseg} A. for samples from the three different scanners that illustrate markedly different image appearance. }
\begin{table}[t]
\centering
\begin{tabular}[width=0.5\textwidth]{c|c|c|c}
        \hline
         \textbf{Scanner}  & \textbf{Field Strength}& \textbf{Voxel Size} (mm)& \textbf{Resolution} \\ \hline
        SA & 1.5 T & $1.2 \times 1 \times 1$ & $128 \times 224 \times 256$    \\ \hline
        PI & 3 T& $1.1 \times 0.5 \times 0.5$ &$144 \times 512 \times 512$   \\ \hline
        SV & 3 T& $0.7 \times 0.7 \times 0.7$ & $261 \times 336 \times 336$ \\ \hline
    \end{tabular}
    \caption{Comparison between MS-SEG scanners.}
    \label{tab:msseg-scanner}
\end{table}

We performed harmonization between each pair of scanners (i.e. SA to SV, SA to PI and PI to SV), where the target domains were selected to be scanners with overall higher image qualities (e.g., higher field strengths, or smaller voxel sizes). 
All images were brain extracted, denoised, and bias field corrected by the challenge organizers. We finally affinely registered them to MNI~\cite{MNI09} coordinates. The groundtruth segmentation \itblue{was} constructed from a consensus of 7 expert-delineated parenchyma lesion masks for each image. 

\begin{table}[t]
    \centering
    \begin{tabular}[width=0.45\textwidth]{c|c|c}
    \hline
     \textbf{Scanner} & Cirrus (CR)
     & Spectralis (SP)
     \\ \hline
     \textbf{Resolution} & $512 \times 1024 \times 128$ & $256 \times 496 \times 49$ \\ \hline
     \textbf{B-scans} & 128 & 49 \\ \hline
     \textbf{Voxel Size}(mm) & $0.01 \times 0.001 \times 0.05$ & $ 0.01 \times 0.004 \times 0.1$\\ \hline
\end{tabular}
\caption{Comparison between OCT scanners.}
\label{tab:retouch_scanner}
\begin{tabular}[width=0.5\textwidth]{c|c|c}
    \hline
     \textbf{Scanner} & GE Healthcare & Philips (PL) \\ \hline
     \textbf{Field Strength} & 1.5T & 3T \\ \hline
     \textbf{Resolution} & $146 \times 256 \times 256$ & $150 \times 256 \times 256$ \\ \hline
     \textbf{Voxel Size}(mm) & $1.2 \times 0.9 \times 0.9$ & $1.2 \times 0.9 \times 0.9$ \\ \hline
\end{tabular}
\caption{Comparison between IXI scanners.}
\label{tab:ixi_scanner}
\end{table}

\subsubsection{RETOUCH} We further tested our method by repurposing the Retinal OCT Fluid Challenge dataset~\cite{RETOUCH}, originally for multi-scanner pathology segmentation. Compared to MRI, distinct OCT scanners show higher imaging variability. The Cirrus (CR) and Spectralis (SP) scanners were treated as source and target respectively, with 24 unpaired scans each. Scanner differences are shown in Table~\ref{tab:retouch_scanner}.

Manual \itblue{expert} segmentation annotations \itblue{were} provided delineating three classes of abnormalities: Intraretinal Fluid (IRF), Subretinal Fluid (SRF) and Pigment Epithelium Detachments (PED). Following~\cite{8759158}, we used nearest-neighbors resampling on CR OCTs to match target dimensionality.

\subsubsection{IXI}
T1w MRI of 241 non-overlapping healthy subjects from two distinct sites were collected from IXI~\cite{IXI}. 18 subject scans with ringing effects or outlier intensity distributions \itblue{were} excluded, \itblue{yielding} 74 scans obtained on a GE Healthcare $1.5$ T scanner (GE) and 167 scans from a Philips $3$ T scanner (PL). Detailed scanner comparisons are given in Table~\ref{tab:ixi_scanner}.

We aim to harmonize GE to PL which displays higher image quality. All T1 scans underwent brain extraction with ROBEX \cite{5742706} \itblue{and bias-field correction to minimize intensity inhomongeneity}. As the dataset does not provide manual segmentation, we simulated a groundtruth \itblue{segmentation} for each image using FSL FAST~\cite{906424} for whole-brain segmentation into three tissue types: grey matter (GM), white matter (WM), and cerebrospinal fluid (CSF). \itblue{Similar pseudo-ground truth estimation approaches have been taken for large-scale medical deep learning} \cite{dalca2018anatomical}.

\subsection{Implementation details:}
The learning rates for the segmentation nets, generators and discriminators \itblue{were} set to $2\times 10^{-4}$, $2\times 10^{-4}$, and $1\times 10^{-4}$, respectively, and Adam optimizers \itblue{were} adopted with $\beta_1 = 0.5$ and $\beta_2 = 0.999$. For IXI, networks \itblue{were} trained on randomly cropped 2D axial slices of size $128\times128$, with batch size 4. For RETOUCH and MS-SEG, the crop size \itblue{was} increased to $256\times 256$, with batch size 2. We empirically set segmentation weights of background/foreground to $0.5/0.5$ for IXI, $0.3/0.7$ for RETOUCH, \itblue{and $0.2/0.8$ for MS-SEG} given the imbalance between positive (lesion) and negative segmentation labels. We set $\lambda_{GAN} =1, \lambda_{cyc}=10, \lambda_{seg}=1$ across datasets.

\subsection{Evaluation scores and baseline methods}
In the absence of paired data, we \itblue{evaluated} harmonization \itblue{performance} from three different perspectives: visual sample fidelity (Sec.~\ref{subsec:fidelity}), downstream task usability (Sec.~\ref{subsec:segment}), as well as sensitivity to translation perturbation and self-adversarial attacks (Sec.~\ref{subsec:attack}). 
We \itblue{conducted} comparisons against the original CycleGAN~\cite{Zhu_2017_ICCV} and two methods which include segmentation-information into cycle-consistent translation: S-CycleGAN~\cite{Zhang_2018_CVPR} and SemGAN~\cite{DBLP:journals/corr/abs-1807-04409}. We used the official CycleGAN implementation from \cite{cyclegan_git}, and reimplemented the other methods as they do not have public code repositories. S-CycleGAN~\cite{Zhang_2018_CVPR} proposed to use additional segmentation networks on the translated outputs, with segmentation loss as an image-level regularization. For consistent comparison, a 2D version of S-CycleGAN was used in our experiments. Adding on to S-CycleGAN, SemGAN~\cite{DBLP:journals/corr/abs-1807-04409} further uses semantic dropout in training by randomly masking out object classes from the inputs to enhance class-to-class translation.
\itblue{As the} MS-lesion masks in MS-SEG are sparse, they cannot be meaningfully dropped out and we excluded comparison with SemGAN on MS-SEG. Sections \ref{subsec:fidelity}, \ref{subsec:segment} and \ref{subsec:attack} provide detailed analysis and comparison. \itblue{All comparisons were conducted on a held-out test set, split at the subject-level.}

\begin{figure*}[t]
\begin{center}
\includegraphics[width=1.0\textwidth]{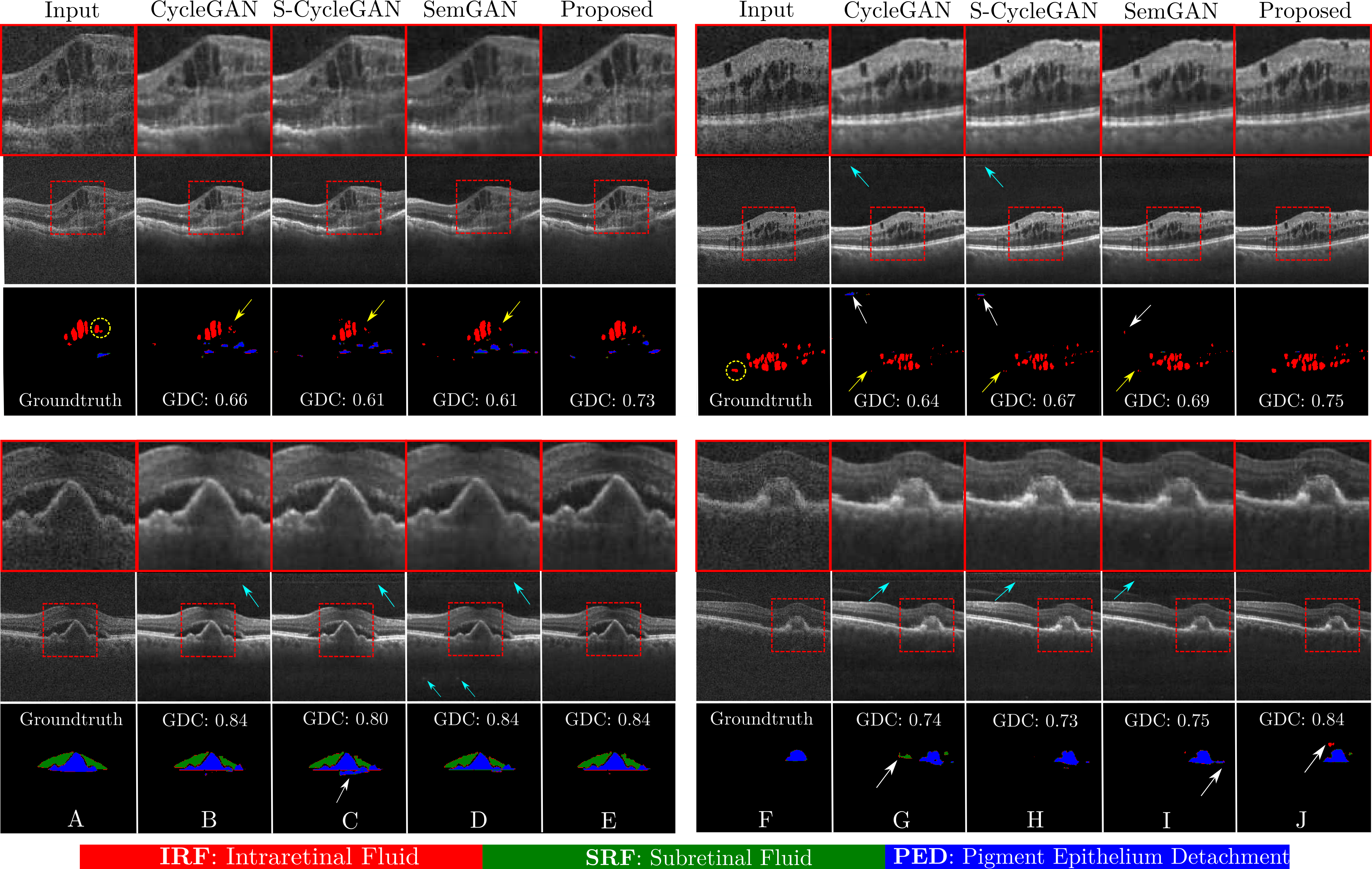}
\end{center}
   \caption{Harmonization results on test multi-site OCT images on RETOUCH. Columns \{\textbf{A, F}\} show the input \itblue{with a zoomed region-of-interest above and} its annotation below; \{\textbf{B, G}\}, \{\textbf{C, H}\}, and \{\textbf{D, I}\} display CycleGAN, S-CycleGAN, SemGAN harmoniz\itblue{ations} and post-hoc segmentations, respectively; \{\textbf{E, J}\} present our results. Generalized Dice Coefficient (GDC) \cite{1717643} is displayed for visualized \itblue{slices}. \itblue{White and yellow arrows show false positive and negative downstream segmentation predictions. Cyan arrows indicate artefacts introduced by translation methods. Readers are encouraged to zoom in for inspection.}} 
\label{fig:result-oct}
\end{figure*}

\begin{figure*}[t]
\begin{center}
\includegraphics[width=1.0\textwidth]{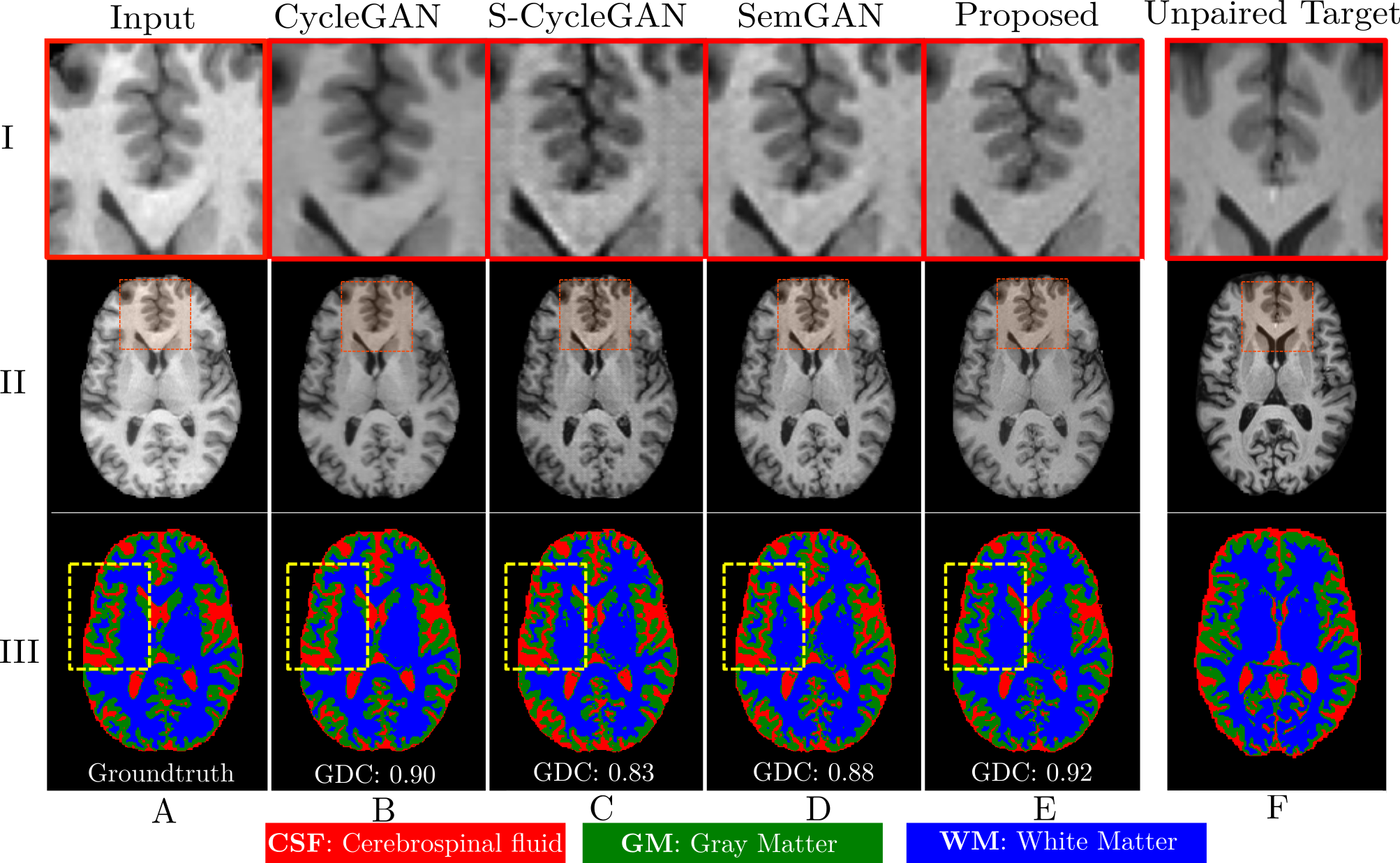}
\end{center}
   \caption{Harmonization results on test multi-site MRI from IXI. Column \textbf{A} shows a T1w axial slice from the source domain with its segmentation groundtruth below; \textbf{B, C, D} show harmonized images from CycleGAN, S-CycleGAN and SemGAN, respectively, \itblue{alongside} the results of a segmentation network trained on the original target domain images; \textbf{E} shows results from our method together with the output from the same segmentation network, revealing an improved post-hoc segmentation (see yellow insets); Generalized Dice Coefficient (GDC)~\cite{1717643} is annotated for each visualized segmentation prediction. \textbf{F} shows an example unpaired target scan. To show finegrained changes, row~\textbf{I} shows a zoomed-in region corresponding to the red \itblue{insets} outlined in row~\textbf{II}. All images were contrast enhanced with gamma correction for improved visualization.}
\label{fig:result-mri}
\end{figure*}

\subsection{Post-harmonization visual fidelity results}
\label{subsec:fidelity}
Representative harmonization results are qualitatively shown in Figs.~\ref{fig:result-msseg},~\ref{fig:result-oct},~\ref{fig:result-mri} for MS-SEG, RETOUCH, and IXI respectively. As observed from the visualizations, baseline methods often produce severe artefacts and lose semantic information crucial to medical image analysis. For example, CycleGAN \itblue{introduced} \itblue{either global} contrast inversion \itblue{(Fig.~\ref{fig:result-msseg} B) or localized contrast inversion to the MS lesion (Fig.~\ref{fig:result-msseg} D). In Fig.~\ref{fig:result-msseg} C, CycleGAN hallucinated artefacts and S-CycleGAN incorrectly removed the MS lesion, whereas our method preserved semantic layout and appearance, with} significantly improved \itblue{downstream} segmentation performance via Dice coefficient. In Fig.~\ref{fig:result-mri} row {I}, we display zoomed-in \itblue{peri}ventricular regions for comparison: S-CycleGAN creates strong checkerboard-like artefacts, whereas our method displays improved image quality in terms of matching the \itblue{fidelity, sharpness,} and contrast of the target domain. \itblue{In OCT harmonization, we observed improved image sharpness and fidelity especially within the red insets of Fig.~\ref{fig:result-oct} as compared to other methods. While we observe no strong artefacts in structured retinal regions across methods, baseline methods hallucinated stripe-like artefacts in the background indicated by cyan arrows in Fig.~\ref{fig:result-oct}, whereas our framework did not.}

\itblue{Further, non-expert human perception may not catch subtle artefacts and distortions present in CNN-synthesized images across datasets which are still detectable via algorithmic means~\cite{Wang_2020_CVPR}. We \itblue{observed} empirical support for this hypothesis as the different translation methods lead to different inception distances and different segmentation results (analyzed below in section~\ref{subsec:segment}) when predicted from the same separately-trained segmentation model, illustrating the underlying differences which affect downstream performance. }

To quantify the visual fidelity of the harmonized results in the absence of paired data, we \itblue{measured} similarity to real unpaired target images in the feature space of a pre-trained network as commonly done with the Fr\'echet and Kernel Inception Distances (FID and KID)~\cite{heusel2017gans,binkowski2018demystifying}. We \itblue{compared} the similarity between source and target domains as a baseline and then between harmonized and target domains. FID measures the similarity of real and synthesized distributions by fitting multivariate Gaussians to feature-space embeddings and calculating the Wasserstein-2 distance between them.
Though commonly used, FID \itblue{has} a strongly biased estimator~\cite{binkowski2018demystifying}, motivating the use of KID which does not assume a parametric distribution on embeddings and applies a polynomial kernel on samples independently drawn from each distribution. For natural images, representations are obtained by running each set of images into an ImageNet-pretrained Inception-v3~\cite{Szegedy_2016_CVPR} network, extracting its last pooling layer features. 

Yet, ImageNet-derived features may not yield good representations for comparing the distributions of embedded medical images. To this end, we \itblue{trained} a multitask autoencoder as the feature extractor to calculate FID and KID. Note that we refer to these scores as FID and KID despite not using Inception-v3 as a feature extractor to maintain consistency with the literature. The feature extractor network is composed of a residual encoder-decoder \itblue{for image reconstruction,} with a domain-classification branch trained on its latent features as in~\cite{le2018supervised}. We \itblue{did} not use skip connections between encoder and decoder to enforce a meaningful bottleneck representation. \itblue{Network configurations and training details are presented in Appendix A.}
As the network both classifies and reconstructs its input, its bottleneck representation is jointly discriminative and reconstructive. Once trained, data from both domains, and the harmonized data are fed into the network for feature extraction.

KID and FID results are presented in Table~\ref{tab:inception}. We see the large domain gap as measured by both KID and FID between source and target domains greatly reduce after harmonization. Compared to baselines, our method achieves significantly improved KID on both MS-SEG and RETOUCH, and comparable KID on IXI. For FID, our method \itblue{showed} consistent improvements on IXI over all compared baselines. We \itblue{skipped} FID on RETOUCH and MS-SEG as FID exhibits strong bias with small sample sizes \cite{binkowski2018demystifying}.

\begin{table*}[h]
\centering
\begin{tabular}{|l|c|c|c|c|c|l|l|c|}
\cline{1-6} \cline{8-9}
\multirow{1}{*}{\multirow{2}{*}{\backslashbox{KID \itblue{($\downarrow$)}}{Dataset}}}       & \multicolumn{3}{c|}{MS-SEG}           & \multirow{2}{*}{RETOUCH} & \multirow{2}{*}{IXI} &  & \multicolumn{1}{l|}{\multirow{2}{*}{\backslashbox{FID \itblue{($\downarrow$)}}{Dataset}}}  & \multirow{2}{*}{IXI} \\ \cline{2-4}
                        & SA $\rightarrow$ SV   & SA $\rightarrow$ PI    & SV $\rightarrow$ PI   &                          &                      &  &                     &                      \\ \cline{1-6} \cline{8-9} 
Source, Target     & \itblue{2.37$\pm$0.17} & \itblue{0.370$\pm$7e-4}  & \itblue{3.09$\pm$1e-2} & 4.65$\pm$1e-6               & 1.140$\pm$4e-3          &  & Source, Target & 385.03              \\
Harmonized, Target~\cite{Zhu_2017_ICCV} & \itblue{1.28$\pm$0.01} & \itblue{0.034$\pm$5e-4} & \itblue{0.18$\pm$3e-3} & 1.81$\pm$1e-6               & \textbf{0.009}$\pm$2e-4          &  & Harmonized, Target~\cite{Zhu_2017_ICCV} & 54.9                 \\
Harmonized, Target~\cite{Zhang_2018_CVPR} & \itblue{2.67$\pm$0.25} & \itblue{\textbf{0.024}$\pm$3e-4} & \itblue{0.87$\pm$2e-2} & 2.56$\pm$1e-6               & 0.013$\pm$4e-4        &  & Harmonized, Target~\cite{Zhang_2018_CVPR} & 4.66                 \\
Harmonized, Target~\cite{DBLP:journals/corr/abs-1807-04409} & -          & -           & -          & 1.87$\pm$1e-6               & 0.014$\pm$5e-4         &  & Harmonized, Target~\cite{DBLP:journals/corr/abs-1807-04409} & 5.02                 \\ 
Harmonized, Target (Ours) & \itblue{\textbf{0.48}$\pm$0.02} & \itblue{\textbf{0.024}$\pm$3e-4} & \itblue{\textbf{0.05}$\pm$3e-3} & \textbf{0.83}$\pm$1e-6               & {0.010}$\pm$2e-4          &  & Harmonized, Target (Ours) & \textbf{3.32}                 \\ \cline{1-6} \cline{8-9} 
\end{tabular}
\caption{Kernel (KID) and Fr\'echet (FID) Inception distances before and after harmonization (lower is better). Given that the FID estimator is strongly biased when the sample size is small~\cite{binkowski2018demystifying}, we skip FID evaluation on RETOUCH and MS-SEG.}
\label{tab:inception}
\end{table*}

\begin{table*}[t]
\centering
\begin{tabular}[t]{l|cc|cc|cc}
\hline
\multirow{2}{*}{\textbf{\itblue{MS-SEG Results (Leave-one-subject-out cross-validation)}}} & \multicolumn{2}{c|}{SA $\rightarrow$ SV} & \multicolumn{2}{c|}{SA $\rightarrow$ PI} & \multicolumn{2}{c}{SV $\rightarrow$ PI} \\ \cline{2-7} 
                                                   & DC \itblue{($\uparrow$)}     &IoU \itblue{$(\uparrow)$}     & DC \itblue{($\uparrow$)}   &IoU \itblue{($\uparrow$)}     & DC \itblue{($\uparrow$)}   &IoU \itblue{($\uparrow$)}  \\ \hline 
Source Segmentor $\rightarrow$ Source Image                        & \itblue{0.53}   & \itblue{0.39}   & \itblue{0.53} & \itblue{0.39}    & \itblue{0.60} & \itblue{0.45}  \\
Target Segmentor $\rightarrow$ Target Image                        & \itblue{0.60}   & \itblue{0.45}    & \itblue{0.57} & \itblue{0.41}    & \itblue{0.57} & \itblue{0.41}\\ \hline
Target Segmentor $\rightarrow$ Source Image                        & \itblue{0.37}   & \itblue{0.26}   & \itblue{0.33}  & \itblue{0.23}  & \itblue{0.43}   &\itblue{0.30}  \\
Target Segmentor $\rightarrow$ Harmonized Image (CycleGAN~\cite{Zhu_2017_ICCV})             & \itblue{0.42}   & \itblue{0.30}  & \itblue{0.31}  & \itblue{0.19}    & \itblue{0.26} & \itblue{0.18}  \\
Target Segmentor $\rightarrow$ Harmonized Image (S-CycleGAN~\cite{Zhang_2018_CVPR})           & \itblue{0.40}   & \itblue{0.29}   &\itblue{0.39} &\itblue{0.28}     & \itblue{0.22}  & \itblue{0.15} \\
Target Segmentor $\rightarrow$ Harmonized Image (Ours)                 & \textbf{\itblue{0.50}} & \textbf{\itblue{0.38}}  & \textbf{\itblue{0.49}}  & \textbf{\itblue{0.35}}  & \textbf{\itblue{0.51}} & \textbf{\itblue{0.37}}\\ \hline \hline
\multirow{2}{*}{\textbf{\itblue{RETOUCH OCT} Results}} & \multicolumn{2}{c|}{IRF} & \multicolumn{2}{c|}{SRF} & \multicolumn{2}{c}{PED} \\ \cline{2-7} 
                                                      & DC \itblue{($\uparrow$)}     &IoU \itblue{($\uparrow$)}   & DC \itblue{($\uparrow$)}    &IoU \itblue{($\uparrow$)}       & DC \itblue{($\uparrow$)}    &IoU \itblue{($\uparrow$)} \\ \hline 
Source Segmentor $\rightarrow$ Source Image                       & 0.71   & 0.55    & 0.54  & 0.42     & 0.55 & 0.40     \\ 
Target Segmentor $\rightarrow$ Target Image                       & 0.72   & 0.56    & 0.70  & 0.54     & 0.11 & 0.06      \\ \hline
Target Segmentor $\rightarrow$ Source Image                       & 0.63   & 0.47    & 0.54  & 0.41     & 0.34 & 0.24      \\ 
Target Segmentor $\rightarrow$ Harmonized Image (CycleGAN~\cite{Zhu_2017_ICCV})            & 0.58   & 0.41    & 0.57   & 0.43    & 0.51 & 0.40      \\ 
Target Segmentor $\rightarrow$ Harmonized Image (S-CycleGAN~\cite{Zhang_2018_CVPR})          & 0.59   & 0.42    & 0.56   & 0.43    & 0.51 & 0.40      \\ 
Target Segmentor $\rightarrow$ Harmonized Image (SemGAN~\cite{DBLP:journals/corr/abs-1807-04409}) & 0.58   & 0.42    & 0.55   & 0.42    & 0.51 & 0.41      \\ 
Target Segmentor $\rightarrow$ Harmonized Image (Proposed)            & \textbf{0.67} & \textbf{0.51}  &\textbf{ 0.62}   & \textbf{0.48}   & \textbf{0.52} & \textbf{0.43}   \\ \hline \hline
\multirow{2}{*}{\textbf{IXI \itblue{MRI} Results}} & \multicolumn{2}{c|}{CSF} & \multicolumn{2}{c|}{GM} & \multicolumn{2}{c}{WM} \\ \cline{2-7}
                                                & DC \itblue{($\uparrow$)}     &IoU \itblue{($\uparrow$)}       & DC \itblue{($\uparrow$)}    & IoU \itblue{($\uparrow$)}    & DC \itblue{($\uparrow$)}      &IoU \itblue{($\uparrow$)}  \\ \hline 
Source Segmentor $\rightarrow$ Source Image                   & 0.90   & 0.83        & 0.91  & 0.83       & 0.88    & 0.81     \\ 
Target Segmentor $\rightarrow$ Target Image                   & 0.90   & 0.83        & 0.89  & 0.82      & 0.88    & 0.81      \\ \hline
Target Segmentor $\rightarrow$ Source Image                     & \textbf{0.85}   & 0.75      & 0.82  & 0.72        & 0.77   & 0.70      \\ 
Target Segmentor $\rightarrow$ Harmonized Image (CycleGAN~\cite{Zhu_2017_ICCV})  & 0.79   & 0.66       & 0.77  & 0.64     & 0.73 & 0.64      \\ 
Target Segmentor $\rightarrow$ Harmonized Image (S-CycleGAN~\cite{Zhang_2018_CVPR})     & 0.61   & 0.46       & 0.64  & 0.48       & 0.71  & 0.61      \\ 
Target Segmentor $\rightarrow$ Harmonized Image (SemGAN~\cite{DBLP:journals/corr/abs-1807-04409}) & 0.80   & 0.68       & 0.76  & 0.64     & 0.72  & 0.64     \\ 
Target Segmentor $\rightarrow$ Harmonized Image (Ours)            & 0.85   & \textbf{0.75}      & \textbf{0.84} & \textbf{0.74} & \textbf{0.81} & \textbf{0.73}    \\    \hline
\end{tabular}
\caption{Post-hoc segmentation results on the 3 settings of the MS-SEG (lesion) dataset, the IXI MRI dataset (CSF: cerebrospinal fluid, GM: grey matter, WM: white matter), and the RETOUCH OCT dataset (IRF: intraretinal fluid, SRF: subretinal fluid, PED: pigment epithelial detachment). \textbf{A}$\rightarrow$\textbf{B} indicates that the segmentation network is trained on domain \textbf{A} and tested on domain \textbf{B}. DC and IoU are Dice coefficient and Intersection over Union, respectively.}
\label{tab: segment_combine}
\end{table*}

\subsection{Post-harmonization segmentation accuracy results} 
\label{subsec:segment}
If an image is correctly translated, then a segmentation network trained on the original target domain images should generalize to the harmonized images. Therefore, we assess the downstream utility of the harmonized images generated by all compared methods by applying domain-specific segmentation networks (trained \textbf{\itblue{separately}} post-hoc on the original domains outside of any translation framework) to the generated harmonized images. Dice Coefficient (DC) and Intersection over Union (IoU) \itblue{were} used as criteria to assess segmentation performance and thus image harmonization.

Figs.~\ref{fig:result-msseg},~\ref{fig:result-oct},~\ref{fig:result-mri} show qualitative comparisons of post-hoc segmentation performance on the harmonized images generated by all methods. Our harmonized images \itblue{were} accurately segmented by the networks trained on the original \itblue{target domain} images, whereas baseline methods \itblue{failed} \itblue{due to sub-optimal translation patterns such as contrast inversion (Fig.~\ref{fig:result-msseg} B, D) and semantic loss (Fig.~\ref{fig:result-msseg} C showing the disappearance of a lesion) and generally result in high segmentation error with false positive (examples marked with white arrows) or false negative predictions (examples marked by yellow arrows).
In Fig.~\ref{fig:result-mri} where labels are densely annotated, we observe minor improvements on the segmentation continuity especially within white matter (see blue labels within yellow insets).}

Quantitatively, DC and IoU are shown in Table \ref{tab: segment_combine} for all datasets.
As baseline \itblue{upper bounds}, we first \itblue{performed} in-domain evaluation (i.e., without harmonization) using two segmentation networks trained on source and target domains, denoted as \textit{`Source Segmentor$\rightarrow$Source Image'} and \textit{`Target Segmentor$\rightarrow$Target Image'}. For MS-SEG, we \itblue{trained} three segmentation networks for the three domains.
We then \itblue{performed} a \itblue{lower bound} cross-domain test by segmenting the source images with the target domain trained network without harmonization \textit{`Target Segmentor$\rightarrow$Source Image'}, observing the expected large performance drop across all scores due to domain shift. On \itblue{using} this target domain trained network to segment harmonized images \textit{`Target Segmentor$\rightarrow$Harmonized Image'}, we expect to see similar performance to \textit{`Source Segmentor$\rightarrow$Source Image'} given that \textit{Harmonized Image} is generated from \textit{Source Image}. 
Harmonized images produced by our method \itblue{achieved} higher quality segmentation in the vast majority of settings evaluated over other baselines, indicating higher downstream utility via smaller domain gap towards the target domain.

\begin{figure*}[t]
\centering
    \begin{subfigure}[t]{\textwidth}
    \includegraphics[width=1.0\textwidth]{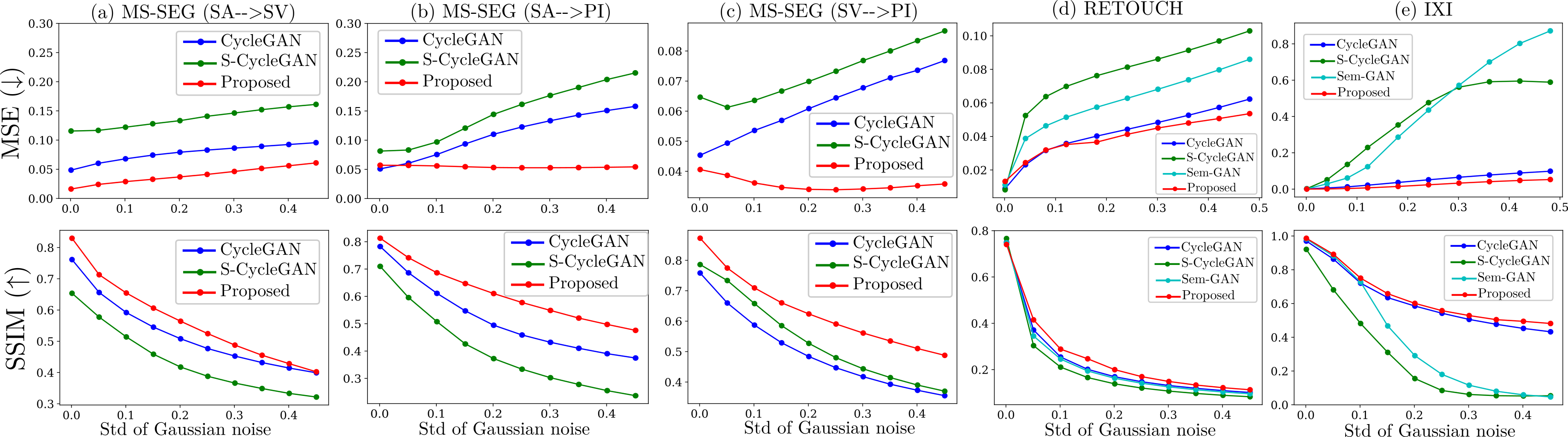}
    \caption{Sensitivity curves of cycle-consistent reconstruction in terms of mean-squared-error (top row, lower is better) and SSIM (bottom row, higher is better) under increasing zero-mean Gaussian perturbation. \itblue{(a) MS-SEG (SA $\rightarrow$ SV), (b) MS-SEG (SA $\rightarrow$ PI), (c) MS-SEG (SV $\rightarrow$ PI)}, (d) RETOUCH, (e) IXI.}    
    \label{fig:sensitivity}
    \end{subfigure}
    \par\bigskip 
    \begin{subfigure}[t]{\textwidth}
    \centering
    \includegraphics[width=1.0\textwidth]{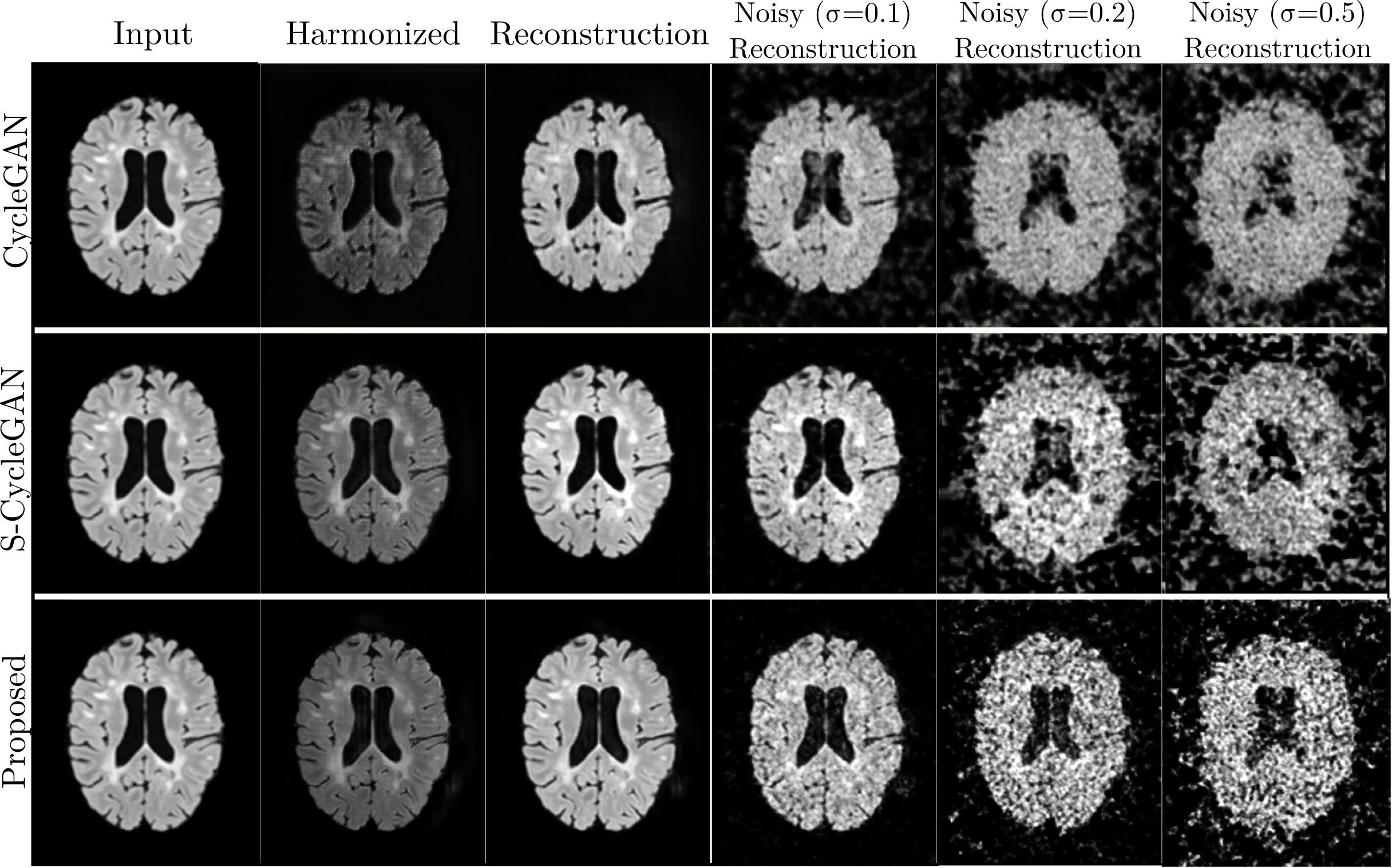}
    \caption{Reconstructions from CycleGAN, S-CycleGAN, and our method on the MS-SEG dataset, under increasing perturbation to the intermediate translation. \itblue{Columns 4--6 show reconstruction results with different levels of Gaussian noise injection with zero mean and standard deviation 0.1, 0.2, and 0.5 applied to the harmonized/translated images}. Our method \itblue{demonstrates} increased robustness to perturbation over competing methods.}
    \label{fig:selfadversarial}
\end{subfigure}
\end{figure*}

\subsection{Sensitivity to self-adversarial attacks}
\label{subsec:attack}
 \itblue{As translation between domains with differing amounts of structural information is ill-posed, a translation GAN trained under cycle-consistency may learn mappings that are highly susceptible to noise as shown in \cite{chu2017cyclegan,bashkirova2019adversarial}, where a complete collapse of cycle-consistent reconstruction is observed when low-amplitude noise is added to the intermediate translation. This susceptibility has been linked to the generator learning to inject human-imperceptible high-frequency noise into the translations as an \textit{adversarial attack} on the discriminator. Therefore, despite the visual appeal of GAN-translations, their adoption for tasks in real-world medical imaging is still limited by these fragile generator mappings which make their translations unreliable for downstream tasks. This problem is exacerbated for medical image harmonization, where the translations are typically further processed by other algorithms for surface extraction, lesion detection, etc.}

\itblue{We speculate that our proposed segmentation-renormalization would empirically improve robustness to such attacks via semantic-regularization.} \itblue{Therefore, we evaluated} the sensitivity $\mathcal{S}$ to self-adversarial attacks of the generator networks by adding zero-mean Gaussian noise with \itblue{increasing} standard deviation to the intermediate translation as in~\cite{bashkirova2019adversarial} and \itblue{measured} the reconstruction error (lower is better) as \itblue{Eq.~\ref{eq:mse}}, where $N$ is the number of 2D slices,
\begin{align}
    \label{eq:mse}
    \mathcal{S}_{\textit{MSE}}(\sigma) =\mathbb{E}_{x \sim X}\left[\|F(G(x) + \mathcal{N}(0, \sigma^2)) - F(G(x))\|_2^2\right].
\end{align}
 We further propose to use the structural similarity index (SSIM) to evaluate reconstruction quality (higher is better), as it correlates better with human perception than MSE \cite{wang2004image},
\begin{align}
\label{eq:ssim}
    \mathcal{S}_{\textit{SSIM}}(\sigma) =\mathbb{E}_{x \sim X}\left[\text{SSIM}(F(G(x) + \mathcal{N}(0, \sigma^2)), F(G(x)))\right].
\end{align}

Fig.~\ref{fig:sensitivity} shows the quantitative effects of linearly increasing $\sigma$ from 0 to 0.5, where we \itblue{found} that the proposed model outperforms the compared models across all datasets in terms of both $\mathcal{S}_{\textit{MSE}}$ and $\mathcal{S}_{\textit{SSIM}}$. Interestingly, we \itblue{found} that S-CycleGAN and Sem-GAN 
are more sensitive to noise than the baseline CycleGAN, \itblue{whereas our approach yields improved SSIM and MSE across all datasets}.

Qualitatively, in Fig.~\ref{fig:selfadversarial} we \itblue{observed} a sharp decline in the perceptual quality of the reconstructions as the noise variance increases. However, the proposed method \itblue{maintained} consistency with its input to a \itblue{perceptually} larger extent as compared to S-CycleGAN and CycleGAN as shown in the rightmost \itblue{three} columns of Fig.~\ref{fig:selfadversarial}. These results illustrate that fine-grained control over the network features via linear scales and shifts based on segmentation empirically makes the translation more robust to self-adversarial attacks \itblue{which may yield improved downstream task performance}.

\section{Discussion}

We present an anatomically-regularized unpaired image-to-image translation framework with a novel segmentation-renormalization. 
\itblue{As quantified by improved KID/FID scores}, our method reduces \itblue{image batch} variability between source and target domains across diverse imaging modalities, while also proving to be \itblue{more} effective for downstream tasks such as structural or lesion segmentation \itblue{as compared to existing translation methods.}
\itblue{Further, as cycle-consistent GANs may produce translations that are corrupted by imperceptible high-frequency self-adversarial noise, we evaluated the sensitivity of all methods to this phenomenon to assess the downstream utility of their translations. We find that our proposed framework outperforms all evaluated baselines,} closing the gap towards reliable real-world medical image translation \itblue{adopted in future studies and biomedical practice.}

Some open issues exist and will be addressed in future work,
\begin{itemize}
    \item \itblue{We assumed stable subject demographics across scanners such that the harmonization experiments focused on imaging batch effects as opposed to biological batch effects.
    However, image harmonization between two different groups (e.g., MRI with scanner A imaging neonatal brains and scanner B imaging pathological adult brains) may be contraindicated. While the proposed model preserves subject-level morphology via individual segmentation information, retaining group-level differences (e.g., age) is not directly addressed. In general, removal of non-biological batch effects while retaining biological batch effects is difficult statistically for even scalar measurements~\cite{nygaard2016methods} and such an extension for image translation GANs will be explored in future work.}
    \item \itblue{The presented framework was developed in 2D for general modality-agnostic applicability. For example, OCT images are highly anisotropic thereby making 3D networks inapplicable. Sequential application of 2D translators to slices from nearly-isotropic volumes (e.g., MRI) may, at times, yield volumetric translations with slice-wise intensity inconsistencies. However, slice consistency can be addressed in a straightforward manner via post-processing extensions such as multi-view fusion~\cite{schilling2019synthesized, Yun2018ImprovementOF} which will be incorporated into future work.}
    \item We use a pseudo-ground truth \itblue{obtained algorithmically~\cite{906424}} in the IXI experiments instead of dense expert annotations which are \itblue{infeasible} to obtain for hundreds of volumes. However, when we do have expert labels \itblue{(RETOUCH and MS-SEG)}, we \itblue{still} observe a strong increase in post-harmonization segmentation performance, indicating \itblue{translation improvements that are relatively insensitive to segmentation quality.} We will explore more effective dense whole-brain segmentation \itblue{approaches} such as multi-atlas methods~\cite{wang2012multi}.
    \item \itblue{We conducted all experiments on public datasets without paired data (i.e. the same subject scanned on two devices). Therefore, qualitative and quantitative comparisons need careful interpretation and analysis. We use surrogate scores of translation quality based on fidelity, distribution matching, segmentation, and robustness in this work. We believe that ideal harmonization validation should include paired held-out subjects scanned in both domains. However, to our knowledge, no such large-scale database of subjects on multiple scanners currently exists.}
\end{itemize}

In summary, our proposed harmonization method improves on previous cycle-consistent adversarial methods, reduces batch effects in multi-center imaging studies, and enables the introduction of large amounts of legacy data into new studies. The presented methodology is fully generic and can be applied in various image translation tasks and is architecture-agnostic, requiring only that the network use normalization layers.

\section{Appendix}

\subsection{Multitask autoencoder for FID/KID}
\begin{figure}[h]
    \centering
    \includegraphics[width=0.2\textwidth]{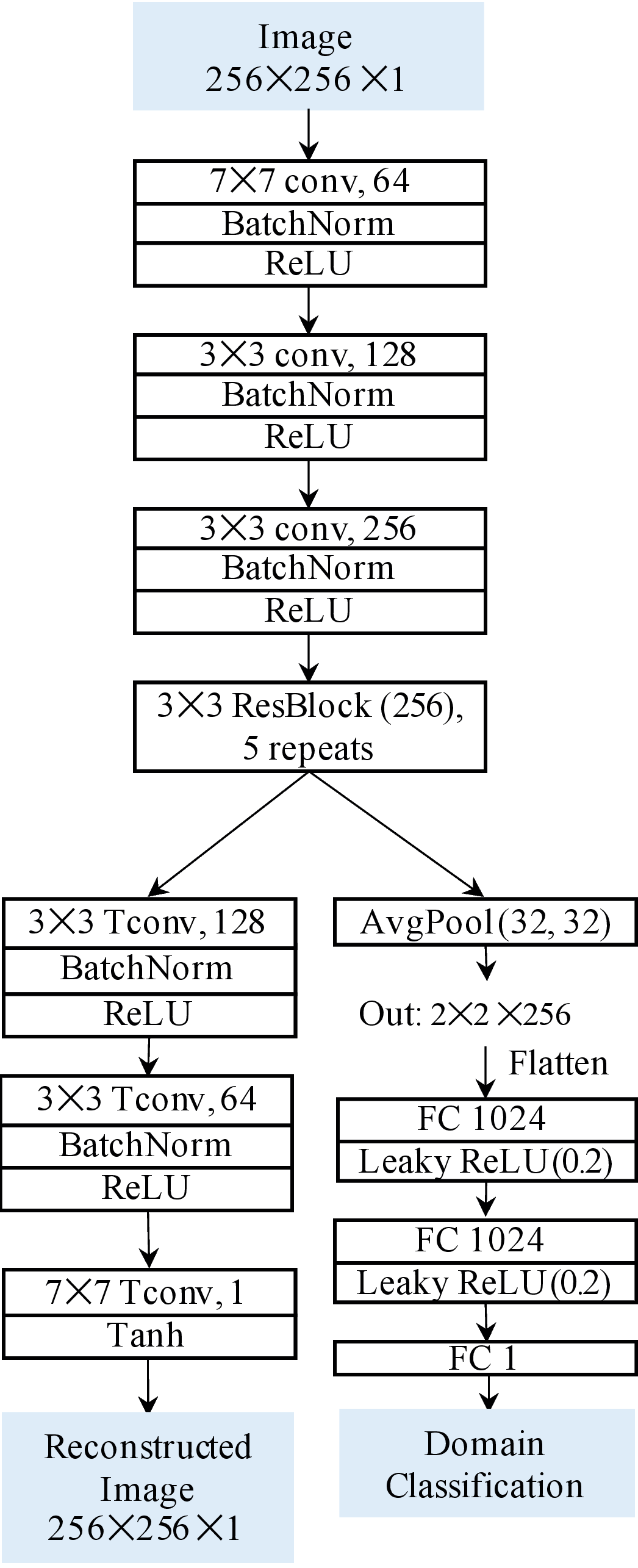}
    \caption{\itblue{Network details of the multitask autoencoder for feature extraction. For IXI, the input/output resolutions are $128\times128\times 1$ and $AvgPool(32,32)$ was replaced by $AvgPool(64,64)$ to accommodate for the output feature size.}}
    \label{fig:net_mae}
\end{figure}

\itblue{The multitask autoencoder described in Fig.~\ref{fig:net_mae} was trained on data from both source and target domains with a binary cross entropy loss for the domain classification branch and an L1 loss for the image reconstruction branch. We used the Adam optimizer with a batch size of 64. Network weights were initialized from $\mathcal{N}(0,0.02)$. The learning rate of the network was set to $0.001$ for the first $20$ epochs and then linearly decayed to zero over the next $100$ epochs.}

\bibliographystyle{IEEEtran}
\bibliography{main}

\begin{thebibliography}{10}
\providecommand{\url}[1]{#1}
\csname url@samestyle\endcsname
\providecommand{\newblock}{\relax}
\providecommand{\bibinfo}[2]{#2}
\providecommand{\BIBentrySTDinterwordspacing}{\spaceskip=0pt\relax}
\providecommand{\BIBentryALTinterwordstretchfactor}{4}
\providecommand{\BIBentryALTinterwordspacing}{\spaceskip=\fontdimen2\font plus
\BIBentryALTinterwordstretchfactor\fontdimen3\font minus
  \fontdimen4\font\relax}
\providecommand{\BIBforeignlanguage}[2]{{%
\expandafter\ifx\csname l@#1\endcsname\relax
\typeout{** WARNING: IEEEtran.bst: No hyphenation pattern has been}%
\typeout{** loaded for the language `#1'. Using the pattern for}%
\typeout{** the default language instead.}%
\else
\language=\csname l@#1\endcsname
\fi
#2}}
\providecommand{\BIBdecl}{\relax}
\BIBdecl

\bibitem{POMPONIO2020116450}
R.~Pomponio \emph{et~al.}, ``Harmonization of large {MRI} datasets for the
  analysis of brain imaging patterns throughout the lifespan,''
  \emph{NeuroImage}, vol. 208, p. 116450, 2020.

\bibitem{GARCIADIAS2020117127}
R.~Garcia-Dias \emph{et~al.}, ``Neuroharmony: A new tool for harmonizing
  volumetric {MRI} data from unseen scanners,'' \emph{NeuroImage}, vol. 220,
  2020.

\bibitem{blumberg2019}
S.~B. Blumberg, M.~Palombo, C.~S. Khoo, C.~M.~W. Tax, R.~Tanno, and D.~C.
  Alexander, ``Multi-stage prediction networks for data harmonization,'' in
  \emph{Medical Image Computing and Computer Assisted Intervention}.\hskip 1em
  plus 0.5em minus 0.4em\relax Cham: Springer International Publishing, 2019,
  pp. 411--419.

\bibitem{10.1093/biostatistics/kxj037}
W.~E. Johnson, C.~Li, and A.~Rabinovic, ``{Adjusting batch effects in
  microarray expression data using empirical Bayes methods},''
  \emph{Biostatistics}, vol.~8, no.~1, pp. 118--127, 04 2006.

\bibitem{FORTIN2018104}
J.-P. Fortin \emph{et~al.}, ``Harmonization of cortical thickness measurements
  across scanners and sites,'' \emph{NeuroImage}, vol. 167, pp. 104 -- 120,
  2018.

\bibitem{10.1007/978-3-030-00536-8_3}
B.~E. Dewey, C.~Zhao, A.~Carass, J.~Oh, P.~A. Calabresi, P.~C.~M. van Zijl, and
  J.~L. Prince, ``Deep harmonization of inconsistent {MR} data for consistent
  volume segmentation,'' in \emph{Simulation and Synthesis in Medical
  Imaging}.\hskip 1em plus 0.5em minus 0.4em\relax Cham: Springer International
  Publishing, 2018, pp. 20--30.

\bibitem{DEWEY2019160}
B.~E. Dewey \emph{et~al.}, ``{DeepHarmony}: A deep learning approach to
  contrast harmonization across scanner changes,'' \emph{Magnetic Resonance
  Imaging}, vol.~64, pp. 160 -- 170, 2019, artificial Intelligence in MRI.

\bibitem{Zhu_2017_ICCV}
J.-Y. Zhu, T.~Park, P.~Isola, and A.~A. Efros, ``Unpaired image-to-image
  translation using cycle-consistent adversarial networks,'' in \emph{The IEEE
  International Conference on Computer Vision (ICCV)}, Oct 2017.

\bibitem{8759158}
P.~{Seeböck} \emph{et~al.}, ``Using cycle{GAN}s for effectively reducing image
  variability across {OCT} devices and improving retinal fluid segmentation,''
  in \emph{IEEE 16th International Symposium on Biomedical Imaging}, 2019.

\bibitem{10.1007/978-3-030-32251-9_52}
F.~Zhao \emph{et~al.}, ``Harmonization of infant cortical thickness using
  surface-to-surface cycle-consistent adversarial networks,'' in \emph{Medical
  Image Computing and Computer Assisted Intervention -- MICCAI}, 2019.

\bibitem{zhang2018harmonic}
R.~Zhang, T.~Pfister, and J.~Li, ``Harmonic {Unpaired} {Image}-to-image
  {Translation},'' in \emph{International Conference on Learning
  Representations}, 2018.

\bibitem{cohen2018distribution}
J.~P. Cohen, M.~Luck, and S.~Honari, ``Distribution matching losses can
  hallucinate features in medical image translation,'' in \emph{International
  conference on medical image computing and computer-assisted
  intervention}.\hskip 1em plus 0.5em minus 0.4em\relax Springer, 2018, pp.
  529--536.

\bibitem{chu2017cyclegan}
C.~Chu, A.~Zhmoginov, and M.~Sandler, ``Cycle{GAN}, a master of
  steganography,'' \emph{arXiv preprint arXiv:1712.02950}, 2017.

\bibitem{bashkirova2019adversarial}
D.~Bashkirova, B.~Usman, and K.~Saenko, ``Adversarial self-defense for
  cycle-consistent {GAN}s,'' \emph{NeurIPS}, pp. 635--645, 2019.

\bibitem{Zhang_2018_CVPR}
Z.~Zhang, L.~Yang, and Y.~Zheng, ``Translating and segmenting multimodal
  medical volumes with cycle- and shape-consistency generative adversarial
  network,'' in \emph{Proceedings of the IEEE Conference on Computer Vision and
  Pattern Recognition (CVPR)}, June 2018.

\bibitem{DBLP:journals/corr/abs-1807-04409}
A.~{Cherian} and A.~{Sullivan}, ``Sem-{GAN}: Semantically-consistent
  image-to-image translation,'' in \emph{IEEE Winter Conference on Applications
  of Computer Vision}, 2019.

\bibitem{mo2018instanceaware}
S.~Mo, M.~Cho, and J.~Shin, ``Instance-aware image-to-image translation,'' in
  \emph{International Conference on Learning Representations}, 2019.

\bibitem{huo2018adversarial}
Y.~Huo, Z.~Xu, S.~Bao, A.~Assad, R.~G. Abramson, and B.~A. Landman,
  ``Adversarial synthesis learning enables segmentation without target modality
  ground truth,'' in \emph{2018 IEEE 15th international symposium on biomedical
  imaging (ISBI 2018)}.\hskip 1em plus 0.5em minus 0.4em\relax IEEE, 2018, pp.
  1217--1220.

\bibitem{chen2020unsupervised}
C.~Chen, Q.~Dou, H.~Chen, J.~Qin, and P.~A. Heng, ``Unsupervised bidirectional
  cross-modality adaptation via deeply synergistic image and feature alignment
  for medical image segmentation,'' \emph{IEEE Transactions on Medical
  Imaging}, 2020.

\bibitem{DBLP:journals/corr/abs-1903-07291}
T.~Park, M.-Y. Liu, T.-C. Wang, and J.-Y. Zhu, ``Semantic image synthesis with
  spatially-adaptive normalization,'' in \emph{Proceedings of the IEEE/CVF
  Conference on Computer Vision and Pattern Recognition}, June 2019.

\bibitem{DBLP:journals/corr/HuangB17}
\BIBentryALTinterwordspacing
X.~Huang and S.~J. Belongie, ``Arbitrary style transfer in real-time with
  adaptive instance normalization,'' \emph{CoRR}, vol. abs/1703.06868, 2017.
  [Online]. Available: \url{http://arxiv.org/abs/1703.06868}
\BIBentrySTDinterwordspacing

\bibitem{perez:hal-01648685}
E.~Perez, F.~Strub, H.~De~Vries, V.~Dumoulin, and A.~Courville, ``{FiLM: Visual
  Reasoning with a General Conditioning Layer},'' in \emph{{AAAI Conference on
  Artificial Intelligence}}, 2018.

\bibitem{kayhan2020translation}
O.~S. Kayhan and J.~C.~v. Gemert, ``On translation invariance in cnns:
  Convolutional layers can exploit absolute spatial location,'' in \emph{IEEE
  Conference on Computer Vision and Pattern Recognition}, 2020.

\bibitem{zhang19shift}
R.~Zhang, ``Making {Convolutional} {Networks} {Shift-Invariant} {Again},'' ser.
  Proceedings of Machine Learning Research, vol.~97.\hskip 1em plus 0.5em minus
  0.4em\relax PMLR, 2019.

\bibitem{He_2016_CVPR}
K.~He, X.~Zhang, S.~Ren, and J.~Sun, ``Deep residual learning for image
  recognition,'' in \emph{The IEEE Conference on Computer Vision and Pattern
  Recognition (CVPR)}, June 2016.

\bibitem{chartsiasdisentangle}
A.~Chartsias, G.~Papanastasiou, S.~Semple, M.~Williams, D.~Newby,
  R.~Dharmakumar, and S.~Tsaftaris, ``Disentangled representation learning in
  cardiac image analysis,'' \emph{Medical Image Analysis}, 2019.

\bibitem{jacenkow2020inside}
G.~Jacenk{\'o}w, A.~Q. O'Neil, B.~Mohr, and S.~A. Tsaftaris, ``Inside: Steering
  spatial attention with non-imaging information in cnns,'' in \emph{Medical
  Image Computing and Computer Assisted Intervention -- MICCAI 2020}.\hskip 1em
  plus 0.5em minus 0.4em\relax Cham: Springer International Publishing, 2020,
  pp. 385--395.

\bibitem{DBLP:journals/corr/RonnebergerFB15}
O.~Ronneberger, P.~Fischer, and T.~Brox, ``U-net: Convolutional networks for
  biomedical image segmentation,'' in \emph{International Conference on Medical
  image computing and computer-assisted intervention}.\hskip 1em plus 0.5em
  minus 0.4em\relax Springer, 2015, pp. 234--241.

\bibitem{rolnick2017deep}
D.~Rolnick, A.~Veit, S.~Belongie, and N.~Shavit, ``Deep learning is robust to
  massive label noise,'' \emph{arXiv preprint arXiv:1705.10694}, 2017.

\bibitem{DBLP:journals/corr/abs-1802-05957}
\BIBentryALTinterwordspacing
T.~Miyato, T.~Kataoka, M.~Koyama, and Y.~Yoshida, ``Spectral {Normalization}
  for {G}enerative {A}dversarial {N}etworks,'' \emph{CoRR}, vol.
  abs/1802.05957, 2018. [Online]. Available:
  \url{http://arxiv.org/abs/1802.05957}
\BIBentrySTDinterwordspacing

\bibitem{mao2017least}
X.~Mao, Q.~Li, H.~Xie, R.~Y. Lau, Z.~Wang, and S.~Paul~Smolley, ``Least squares
  generative adversarial networks,'' in \emph{Proceedings of the IEEE
  international conference on computer vision}, 2017, pp. 2794--2802.

\bibitem{IXI}
``{IXI} brain database,'' \url{http://brain-development.org/ixi-dataset/},
  accessed: 2020-03-14.

\bibitem{commowick_objective_2018}
O.~Commowick \emph{et~al.}, ``\BIBforeignlanguage{en}{Objective {Evaluation} of
  {Multiple} {Sclerosis} {Lesion} {Segmentation} using a {Data} {Management}
  and {Processing} {Infrastructure}},''
  \emph{\BIBforeignlanguage{en}{Scientific Reports}}, vol.~8, no.~1, p. 13650,
  Dec. 2018.

\bibitem{RETOUCH}
H.~Bogunovi\'c \emph{et~al.}, ``{RETOUCH - The Retinal OCT Fluid Detection and
  Segmentation Benchmark and Challenge},'' \emph{IEEE Transactions on Medical
  Imaging}, vol.~38, no.~8, pp. 1858--1874, Aug 2019.

\bibitem{MNI09}
V.~Fonov, A.~Evans, R.~Mckinstry, C.~Almli, and L.~Collins, ``Unbiased
  nonlinear average age-appropriate brain templates from birth to adulthood,''
  \emph{Neuroimage}, vol.~47, 07 2009.

\bibitem{5742706}
J.~E. {Iglesias}, C.~{Liu}, P.~M. {Thompson}, and Z.~{Tu}, ``Robust brain
  extraction across datasets and comparison with publicly available methods,''
  \emph{IEEE Transactions on Medical Imaging}, vol.~30, no.~9, pp. 1617--1634,
  Sep. 2011.

\bibitem{906424}
Y.~{Zhang}, M.~{Brady}, and S.~{Smith}, ``Segmentation of brain {MR} images
  through a hidden markov random field model and the expectation-maximization
  algorithm,'' \emph{IEEE Transactions on Medical Imaging}, vol.~20, no.~1, pp.
  45--57, Jan 2001.

\bibitem{dalca2018anatomical}
A.~V. Dalca, J.~Guttag, and M.~R. Sabuncu, ``Anatomical priors in convolutional
  networks for unsupervised biomedical segmentation,'' in \emph{Proceedings of
  the IEEE Conference on Computer Vision and Pattern Recognition}, 2018, pp.
  9290--9299.

\bibitem{cyclegan_git}
``Cycle{GAN} {I}mplementation,''
  \url{https://github.com/junyanz/pytorch-CycleGAN-and-pix2pix}, accessed:
  2020-08-19.

\bibitem{1717643}
W.~R. {Crum}, O.~{Camara}, and D.~L.~G. {Hill}, ``Generalized overlap measures
  for evaluation and validation in medical image analysis,'' \emph{IEEE
  Transactions on Medical Imaging}, vol.~25, no.~11, pp. 1451--1461, 2006.

\bibitem{Wang_2020_CVPR}
S.-Y. Wang, O.~Wang, R.~Zhang, A.~Owens, and A.~A. Efros, ``Cnn-generated
  images are surprisingly easy to spot... for now,'' in \emph{IEEE/CVF
  Conference on Computer Vision and Pattern Recognition (CVPR)}, June 2020.

\bibitem{heusel2017gans}
M.~Heusel, H.~Ramsauer, T.~Unterthiner, B.~Nessler, and S.~Hochreiter, ``{GAN}s
  trained by a two time-scale update rule converge to a local nash
  equilibrium,'' 2017.

\bibitem{binkowski2018demystifying}
M.~Bińkowski, D.~J. Sutherland, M.~Arbel, and A.~Gretton, ``Demystifying {MMD}
  {GAN}s,'' in \emph{International Conference on Learning Representations},
  2018.

\bibitem{Szegedy_2016_CVPR}
C.~Szegedy, V.~Vanhoucke, S.~Ioffe, J.~Shlens, and Z.~Wojna, ``Rethinking the
  inception architecture for computer vision,'' in \emph{Proceedings of the
  IEEE Conference on Computer Vision and Pattern Recognition}, 2016.

\bibitem{le2018supervised}
L.~Le, A.~Patterson, and M.~White, ``Supervised autoencoders: Improving
  generalization performance with unsupervised regularizers,'' in
  \emph{Advances in Neural Information Processing Systems}, 2018, pp. 107--117.

\bibitem{wang2004image}
Z.~Wang, A.~C. Bovik, H.~R. Sheikh, and E.~P. Simoncelli, ``Image quality
  assessment: from error visibility to structural similarity,'' \emph{IEEE
  transactions on image processing}, vol.~13, no.~4, pp. 600--612, 2004.

\bibitem{nygaard2016methods}
V.~Nygaard, E.~A. R{\o}dland, and E.~Hovig, ``Methods that remove batch effects
  while retaining group differences may lead to exaggerated confidence in
  downstream analyses,'' \emph{Biostatistics}, 2016.

\bibitem{schilling2019synthesized}
K.~G. Schilling, J.~Blaber, Y.~Huo, A.~Newton, C.~Hansen, V.~Nath, A.~T.
  Shafer, O.~Williams, S.~M. Resnick, B.~Rogers \emph{et~al.}, ``Synthesized b0
  for diffusion distortion correction (synb0-disco),'' \emph{Magnetic resonance
  imaging}, vol.~64, pp. 62--70, 2019.

\bibitem{Yun2018ImprovementOF}
J.~Yun, M.~Lee, H.~Park, J.~Lee, J.~Seo, and Namkug, ``Improvement of fully
  automated airway segmentation on computed tomographic images using 2.5 d and
  3d convolutional neural net,'' \emph{Medical Image Analysis}, vol.~51, pp.
  13--20, 01 2019.

\bibitem{wang2012multi}
H.~Wang, J.~W. Suh, S.~R. Das, J.~B. Pluta, C.~Craige, and P.~A. Yushkevich,
  ``Multi-atlas segmentation with joint label fusion,'' \emph{IEEE transactions
  on pattern analysis and machine intelligence}, vol.~35, no.~3, pp. 611--623,
  2012.

\end{thebibliography}

\end{document}